%% file: main.tex
\documentclass[%
 reprint,
 unsortedaddress,
superscriptaddress,
nofootinbib,
 amsmath,amssymb, amsfonts,
 aps,
 prd,
floatfix,
showkeys,twoside
]{revtex4-2}

\renewcommand{\selectlanguage}[1]{}

\usepackage{graphicx}% Include figure files
\usepackage{dcolumn}% Align table columns on decimal point
\usepackage[mathlines, switch]{lineno}% Enable numbering of text and display math
\usepackage{bm}% bold math
\usepackage[hidelinks]{hyperref}% add hypertext capabilities
\usepackage[nameinlink,capitalize]{cleveref}
\usepackage{xspace}
\usepackage{amsthm}
\usepackage{mathtools}
\usepackage{physics}
\usepackage[dvipsnames]{xcolor}
\usepackage{adjustbox}
\usepackage{placeins}
\usepackage[T1]{fontenc}
\usepackage{csquotes}
\usepackage{booktabs}
\usepackage{array}
\usepackage{fnpct} % spacing around footnotes

\usepackage{xargs}
\usepackage{soul}
\usepackage[normalem]{ulem}
\usepackage{cancel}

\usepackage{ifthen}
\usepackage{acro3p7}

\DeclareMathOperator\erfc{erfc}

\newboolean{showunsure}
\newboolean{showchange}
\newboolean{showinfo}
\newboolean{showimprovement}
\newboolean{showadd}
\newboolean{shownote}
\newboolean{showresponse}

\newboolean{hidescratch}

\setboolean{showimprovement}{true}
\setboolean{showchange}{true}
\setboolean{showadd}{true}
\setboolean{showunsure}{false}
\setboolean{showresponse}{true}

\setboolean{hidescratch}{true}

\ifthenelse{\boolean{hidescratch}}
  {\setboolean{showinfo}{false}}
  {\setboolean{showinfo}{true}}
\ifthenelse{\boolean{hidescratch}}
  {\setboolean{shownote}{false}}
  {\setboolean{shownote}{true}}
\usepackage[colorinlistoftodos,prependcaption,textsize=tiny, textwidth=1.5cm]{todonotes}
\newlength{\marginshift}
\newsavebox{\leftmarginbox}
\newsavebox{\rightmarginbox}

\newcommand{\shiftedmarginpar}[2][]{%
  \savebox\leftmarginbox{\parbox{\marginparwidth}{#1}}%
  \savebox\rightmarginbox{\parbox{\marginparwidth}{#2}}%
  \ifvoid\leftmarginbox\else \ifdim\marginshift>\dp\leftmarginbox \marginshift=\dp\leftmarginbox\fi\fi
  \ifvoid\rightmarginbox\else \ifdim\marginshift>\dp\rightmarginbox \marginshift=\dp\rightmarginbox\fi\fi
  \marginpar[\ifvoid\leftmarginbox\else \raisebox{\marginshift}{\usebox\leftmarginbox}\fi]%
    {\ifvoid\rightmarginbox\else \raisebox{\marginshift}{\usebox\rightmarginbox}\fi}%
  \setlength{\marginshift}{0pt}}
  
\makeatletter
 \patchcmd{\@todonotes@drawMarginNoteWithLine}{\marginpar}{\shiftedmarginpar}{}{failed}
\makeatother
\newcommandx{\unsure}[2][1=]{%
    \ifthenelse{\boolean{showunsure}}%
        {\todo[linecolor=red,backgroundcolor=red!25,bordercolor=red,#1]{#2}}%
        {\todo[disable,#1]{#2}}%
    }%
\newcommandx{\change}[2][1=]{%
    \ifthenelse{\boolean{showchange}}%
        {\todo[linecolor=blue,backgroundcolor=blue!25,bordercolor=blue,#1]{#2}}%
        {\todo[disable,#1]{#2}}%
    }%
\newcommandx{\info}[2][1=]{%
    \ifthenelse{\boolean{showinfo}}%
        {\todo[linecolor=OliveGreen,backgroundcolor=OliveGreen!25,bordercolor=OliveGreen, #1]{#2}}%
        {\todo[disable,#1]{#2}}%
    }%
\newcommandx{\response}[2][1=]{%
    \ifthenelse{\boolean{showresponse}}%
        {\todo[linecolor=Aquamarine,backgroundcolor=Aquamarine!25,bordercolor=Aquamarine, #1]{#2}}%
        {\todo[disable,#1]{#2}}%
    }%
\newcommandx{\improvement}[2][1=]{%
    \ifthenelse{\boolean{showimprovement}}%
        {\todo[linecolor=Plum,backgroundcolor=Plum!25,bordercolor=Plum,#1]{#2}}%
        {\todo[disable,#1]{#2}}%
    }%
\newcommandx{\add}[2][1=]{%
    \ifthenelse{\boolean{showadd}}%
        {\todo[linecolor=Goldenrod,backgroundcolor=Goldenrod!25,bordercolor=Goldenrod,#1]{#2}}%
        {\todo[disable,#1]{#2}}%
    }%
\newcommandx{\note}[2][1=]{
    \ifthenelse{\boolean{shownote}}
        {\todo[linecolor=Emerald,backgroundcolor=Emerald!25,bordercolor=Emerald,#1]{#2}}%
        {\todo[disable,#1]{#2}}%
    }%
\newcommandx{\hiddentodo}[2][1=]{\todo[disable,#1]{#2}}%

\newboolean{isthesis}
\setboolean{isthesis}{false}

\acsetup{
    make-links, 
    list/exclude={unit,common,convenience},
    cite/group = true,
    }
\input{acro_local.tex}

\definecolor{redc}{RGB}{138, 22, 3}

\definecolor{blue-violet}{rgb}{0.54, 0.17, 0.89}
\definecolor{black}{rgb}{0.0, 0.0, 0.0}
\newcommand*{\Colin}{\textcolor{blue-violet}}
\definecolor{carminered}{rgb}{1.0, 0.0, 0.22}

\definecolor{amber}{rgb}{0.95, 0.8, 0.2}

\definecolor{pink}{RGB}{255, 20, 147}     

\definecolor{byzantine}{rgb}{0.74, 0.2, 0.64}

\definecolor{comment}{RGB}{166, 38, 164}

\newcommand*{\rev}{\textcolor{black}}

\def\kms{\ifmmode{~{\rm km~s^{-1}}}\else{~km s$^{-1}$}\fi}
\def\cm3{\ifmmode{~{\rm cm^{-3}}}\else{~cm$^{-3}$}\fi}
\def\Ms{\ifmmode{~{\rm M_\odot}}\else{M$_\odot$}\fi}
\def\gunit{\ensuremath{{\rm~\times~10^{-11}~GeV^{-1}}}\xspace}
\def\neV{\ensuremath{\rm~neV}\xspace}
\def\eV{\ensuremath{\rm~eV}\xspace}
\newcommand{\alpphot}{\acs*{alp}-photon\xspace}
\newcommand{\pkg}[1]{\texttt{#1}} % textsc?

\def\ltsima{$\; \buildrel < \over \sim \;$}
\def\simlt{\lower.5ex\hbox{\ltsima}} % < over ~
\def\gtsima{$\; \buildrel > \over \sim \;$}
\def\simgt{\lower.5ex\hbox{\gtsima}} % > over ~

\def\gray{$\gamma$-ray\xspace}

\def\graysnoh{$\gamma$~rays\xspace}
\def\TS{\ifmmode{\mathrm{TS}}\else{TS}\xspace\fi}
\def\Lik{\mathcal{L}}
\newcommand{\axm}{\ensuremath{m_a}\xspace}
\newcommand{\axmhat}{\ensuremath{\hat{m}_a}\xspace}
\newcommand{\axg}{\ensuremath{g_{a\gamma}}\xspace}\newcommand{\axghat}{\ensuremath{\hat{g}_{a\gamma}}\xspace}
\def\Fermi{\textit{Fermi}\xspace}

\makeatletter
\renewcommand*\env@matrix[1][\arraystretch]{%
  \edef\arraystretch{#1}%
  \hskip -\arraycolsep
  \let\@ifnextchar\new@ifnextchar
  \array{*\c@MaxMatrixCols c}}
\makeatother
\begin{document}
% \linenumbers\relax % Commence numbering lines

\title{Constraints on Axion-Like Particles from VERITAS Observations of a Flaring Radio Galaxy in the Perseus Cluster}

\author{C.~B.~Adams}
\email[Email address: ]{ca2762@columbia.edu}
\affiliation{Physics Department, Columbia University, New York, NY 10027, USA}
\author{A.~Archer}\affiliation{Department of Physics and Astronomy, DePauw University, Greencastle, IN 46135-0037, USA}
\author{P.~Bangale}\affiliation{Department of Physics and Astronomy and the Bartol Research Institute, University of Delaware, Newark, DE 19716, USA}
\author{J.~T.~Bartkoske}\affiliation{Department of Physics and Astronomy, University of Utah, Salt Lake City, UT 84112, USA}
\author{W.~Benbow}\affiliation{Center for Astrophysics $|$ Harvard \& Smithsonian, Cambridge, MA 02138, USA}
\author{Y.~Chen}\affiliation{Department of Physics and Astronomy, University of California, Los Angeles, CA 90095, USA}
\author{J.~L.~Christiansen}\affiliation{Physics Department, California Polytechnic State University, San Luis Obispo, CA 94307, USA}
\author{A.~J.~Chromey}\affiliation{Center for Astrophysics $|$ Harvard \& Smithsonian, Cambridge, MA 02138, USA}
\author{A.~Duerr}\affiliation{Department of Physics and Astronomy, University of Utah, Salt Lake City, UT 84112, USA}
\author{M.~Errando}\affiliation{Department of Physics, Washington University, St. Louis, MO 63130, USA}
\author{M.~Escobar~Godoy}\affiliation{Santa Cruz Institute for Particle Physics and Department of Physics, University of California, Santa Cruz, CA 95064, USA}
\author{J.~Escudero~Pedrosa}\affiliation{Harvard-Smithsonian Center for Astrophysics, 60 Garden Street, Cambridge, MA 02138, USA}
\author{S.~Feldman}\affiliation{Department of Physics and Astronomy, University of California, Los Angeles, CA 90095, USA}
\author{Q.~Feng}\email[Email address: ]{qi.feng@utah.edu}
\affiliation{Department of Physics and Astronomy, University of Utah, Salt Lake City, UT 84112, USA}
\author{S.~Filbert}\affiliation{Department of Physics and Astronomy, University of Utah, Salt Lake City, UT 84112, USA}
\author{L.~Fortson}\affiliation{School of Physics and Astronomy, University of Minnesota, Minneapolis, MN 55455, USA}
\author{A.~Furniss}\affiliation{Santa Cruz Institute for Particle Physics and Department of Physics, University of California, Santa Cruz, CA 95064, USA}
\author{W.~Hanlon}\affiliation{Center for Astrophysics $|$ Harvard \& Smithsonian, Cambridge, MA 02138, USA}
\author{O.~Hervet}\affiliation{Santa Cruz Institute for Particle Physics and Department of Physics, University of California, Santa Cruz, CA 95064, USA}
\author{C.~E.~Hinrichs}\affiliation{Center for Astrophysics $|$ Harvard \& Smithsonian, Cambridge, MA 02138, USA and Department of Physics and Astronomy, Dartmouth College, 6127 Wilder Laboratory, Hanover, NH 03755 USA}
\author{J.~Holder}\affiliation{Department of Physics and Astronomy and the Bartol Research Institute, University of Delaware, Newark, DE 19716, USA}
\author{Z.~Hughes}\affiliation{Department of Physics, Washington University, St. Louis, MO 63130, USA}
\author{T.~B.~Humensky}\affiliation{Department of Physics, University of Maryland, College Park, MD, USA and NASA GSFC, Greenbelt, MD 20771, USA}
\author{M.~Iskakova}\affiliation{Department of Physics, Washington University, St. Louis, MO 63130, USA}
\author{W.~Jin}\affiliation{Department of Physics and Astronomy, University of California, Los Angeles, CA 90095, USA}
\author{P.~Kaaret}\affiliation{Department of Physics and Astronomy, University of Iowa, Van Allen Hall, Iowa City, IA 52242, USA}
\author{M.~Kertzman}\affiliation{Department of Physics and Astronomy, DePauw University, Greencastle, IN 46135-0037, USA}
\author{M.~Kherlakian}\affiliation{Fakult\"at f\"ur Physik \& Astronomie, Ruhr-Universit\"at Bochum, D-44780 Bochum, Germany}
\author{D.~Kieda}\affiliation{Department of Physics and Astronomy, University of Utah, Salt Lake City, UT 84112, USA}
\author{T.~K.~Kleiner}\affiliation{DESY, Platanenallee 6, 15738 Zeuthen, Germany}
\author{N.~Korzoun}\affiliation{Department of Physics and Astronomy and the Bartol Research Institute, University of Delaware, Newark, DE 19716, USA}
\author{F.~Krennrich}\affiliation{Department of Physics and Astronomy, Iowa State University, Ames, IA 50011, USA}
\author{S.~Kumar}\affiliation{Department of Physics, University of Maryland, College Park, MD, USA }
\author{S.~Kundu}\affiliation{Department of Physics and Astronomy, University of Alabama, Tuscaloosa, AL 35487, USA}
\author{M.~J.~Lang}\affiliation{School of Natural Sciences, University of Galway, University Road, Galway, H91 TK33, Ireland}
\author{M.~Lundy}\affiliation{Physics Department, McGill University, Montreal, QC H3A 2T8, Canada}
\author{P.~Moriarty}\affiliation{School of Natural Sciences, University of Galway, University Road, Galway, H91 TK33, Ireland}
\author{R.~Mukherjee}\affiliation{Department of Physics and Astronomy, Barnard College, Columbia University, NY 10027, USA}
\author{W.~Ning}\affiliation{Department of Physics and Astronomy, University of California, Los Angeles, CA 90095, USA}
\author{R.~A.~Ong}\affiliation{Department of Physics and Astronomy, University of California, Los Angeles, CA 90095, USA}
\author{A.~Pandey}\affiliation{Department of Physics and Astronomy, University of Utah, Salt Lake City, UT 84112, USA}
\author{M.~Pohl}\affiliation{Institute of Physics and Astronomy, University of Potsdam, 14476 Potsdam-Golm, Germany and DESY, Platanenallee 6, 15738 Zeuthen, Germany}
\author{E.~Pueschel}\affiliation{Fakult\"at f\"ur Physik \& Astronomie, Ruhr-Universit\"at Bochum, D-44780 Bochum, Germany}
\author{J.~Quinn}\affiliation{School of Physics, University College Dublin, Belfield, Dublin 4, Ireland}
\author{P.~L.~Rabinowitz}\affiliation{Department of Physics, Washington University, St. Louis, MO 63130, USA}
\author{K.~Ragan}\affiliation{Physics Department, McGill University, Montreal, QC H3A 2T8, Canada}
\author{P.~T.~Reynolds}\affiliation{Department of Physical Sciences, Munster Technological University, Bishopstown, Cork, T12 P928, Ireland}
\author{D.~Ribeiro}\affiliation{School of Physics and Astronomy, University of Minnesota, Minneapolis, MN 55455, USA}
\author{E.~Roache}\affiliation{Center for Astrophysics $|$ Harvard \& Smithsonian, Cambridge, MA 02138, USA}
\author{C.~Rulten}\affiliation{School of Physics and Astronomy, University of Minnesota, Minneapolis, MN 55455, USA}\affiliation{Centre for Advanced Instrumentation, Department of Physics, University of Durham, South Road, Durham DH1 3LE, UK}
\author{I.~Sadeh}\affiliation{DESY, Platanenallee 6, 15738 Zeuthen, Germany}
\author{L.~Saha}\affiliation{Center for Astrophysics $|$ Harvard \& Smithsonian, Cambridge, MA 02138, USA}
\author{M.~Santander}\affiliation{Department of Physics and Astronomy, University of Alabama, Tuscaloosa, AL 35487, USA}
\author{G.~H.~Sembroski}\affiliation{Department of Physics and Astronomy, Purdue University, West Lafayette, IN 47907, USA}
\author{R.~Shang}\affiliation{Department of Physics and Astronomy, Barnard College, Columbia University, NY 10027, USA}
\author{D.~Tak}\affiliation{SNU Astronomy Research Center, Seoul National University, Seoul 08826, Republic of Korea.}
\author{A.~K.~Talluri}\affiliation{School of Physics and Astronomy, University of Minnesota, Minneapolis, MN 55455, USA}
\author{J.~V.~Tucci}\affiliation{Department of Physics, Indiana University Indianapolis, Indianapolis, Indiana 46202, USA}
\author{J.~Valverde}\affiliation{Department of Physics, University of Maryland, Baltimore County, Baltimore MD 21250, USA and NASA GSFC, Greenbelt, MD 20771, USA}
\author{V.~V.~Vassiliev}\affiliation{Department of Physics and Astronomy, University of California, Los Angeles, CA 90095, USA}
\author{D.~A.~Williams}\affiliation{Santa Cruz Institute for Particle Physics and Department of Physics, University of California, Santa Cruz, CA 95064, USA}
\author{S.~L.~Wong}\affiliation{Physics Department, McGill University, Montreal, QC H3A 2T8, Canada}
\author{J.~Woo}\affiliation{Columbia Astrophysics Laboratory, Columbia University, New York, NY 10027, USA}
\author{T.~Yoshikoshi}\affiliation{Institute for Cosmic Ray Research, University of Tokyo, 5-1-5, Kashiwa-no-ha, Kashiwa, Chiba 277-8582, Japan}

\collaboration{The VERITAS Collaboration}
\noaffiliation

\author{M.~Meyer}
\affiliation{CP3-Origins, University of Southern Denmark, Campusvej 55, 5230 Odense M, Denmark}

\date{\today}% It is always \today, today,

\begin{abstract}

\begin{description}
\item[Background]
\Acp*{alp} are hypothetical particles that emerge in numerous theoretical extensions to the Standard Model. Their coupling to electromagnetic field implies that ALPs would mix with photons in the presence of external magnetic fields. As ALP phenomenology is governed by the mass and strength of its coupling, there is a subset of this parameter space in which this mixing would be expected to leave an imprint on the spectra of TeV \gray sources.
\item[Data] In 2017, the \acs*{vts} \gray observatory recorded the second day of a dramatic flare of the \acl*{rg} NGC~1275, embedded at the center of the Perseus galaxy cluster. This serendipitous locale provides a spatially-extended magnetic field of strength $\mathcal{O}{(10{\rm~\mu G})}$ through which escaping photons traverse, making it an excellent target to study ALPs.
\item[Methods] We analyze the \acs*{vts} data of NGC~1275's 2017 flare with the \pkg{gammapy} analysis package. Extensive fitting and modeling are performed to ultimately conduct a likelihood analysis used to search for any evidence of a preference for ALPs and to explore the confidence with which constraints can be set. We adopt the ${\rm CL_s}$ method for this study for its conservative approach to setting limits in regimes where the search has limited sensitivity.
\item[Results] %We document no evidence for the existence of \acp*{alp}. 
No evidence for the existence of ALPs is found, and no combination of mass and coupling strength can be excluded at or above 95\% confidence level. We provide a map showing the strength of our exclusions in the mass and coupling parameter space. The strongest exclusions are found in the mass range ${2 \times 10^{-7} {\rm~eV} \lesssim \axm \lesssim 4 \times 10^{-7} {\rm~eV}}$ and at the coupling strength of ${\axg \gtrsim 3 \gunit}$ up to 80\% confidence level, which are consistent with previous studies. %Portions of the mass and coupling parameter space are excluded at up to the 80\% confidence level.
\item[Conclusions] We find the ${\rm CL_s}$ to be a trustworthy approach, and advocate for its continued usage in future studies. We note that many of the limitations contributing to the limited sensitivity seen by \acs*{vts} in this study will be improved with next-generation \gray instruments, such as the \ac*{ctao}.
\end{description}
\end{abstract}

\maketitle

% \tableofcontents

%%% Below is a fix for weird table of contents formatting %%%
\makeatletter
\let\toc@pre\relax
\let\toc@post\relax
\makeatother
%%%%%%%%%%%%%%%%%%%%%%%%%%%%%%%%%%%%%%%%%%%%%%%%%%%%%%%%%%%%%

% Can use \textbackslash emph to \emph{emphasize} and \textbackslash [Your Name] to add a contribution identifiable to you by its \Colin{color}. \textbackslash citet\{\} for in-line citations and \textbackslash citep\{\} for bracketed citations.

% I’ve also included some functionality to add TODOs/comments in the form of the following commands:
% \begin{itemize}
%     \item \textbackslash unsure\{Use if not sure about something\}
%     \item \textbackslash change\{Use to indicate a change needs to be made\}
%     \item \textbackslash info\{Use to provide info\}
%     \item \textbackslash improvement\{Use to indicate an improvement that can be made\}
%     \item \textbackslash add\{Use to indicate something that needs to be added\}
%     \item \textbackslash note\{Use to add a general note\}
%     \item \textbackslash missingfigure\{Use as a placeholder for a planned figure\}
% \end{itemize}

% All of the commmands above, except for the last one, can be supplied with the option [inline] so that the TODO shows up in the text as opposed to in the margin. For example: \textbackslash add[inline]\{add text about the TeV emission\}

% \listoftodos[To Dos and Comments]

\section{Introduction}
\label{sec:alpintro}
TeV \gray observations can be used to probe fundamental physics beyond the \ac{sm} of particle physics \citep[see, e.g.,][]{the_cta_consortium_science_2019}, including axions and \acp{alp} \citep[see, e.g.,][and references therein]{abdalla_sensitivity_2021, choi_recent_2021}. 
Axions are hypothetical \ac{png} bosons postulated to solve the ``strong \ac{cp} problem'' in particle physics and, fortuitously, are excellent candidates for \ac{dm} \citep[see, e.g.,][]{chadha-day_axion_2022}. Axions would couple with photon pairs, with a coupling strength proportional to their mass, which is unconstrained in theoretical models. \Acp{alp}, on the other hand, are predicted in many extensions of the \ac{sm}, particularly in numerous theories of string compactifications, with a looser constraint: their coupling strength to photons $\axg$ is independent of mass $\axm$ \citep[e.g.,][]{kim_light_1987, turok_almost-goldstone_1996, jaeckel_low-energy_2010, ringwald_searching_2014, dias_quest_2014, acharya_m_2010, svrcek_axions_2006, choi_accions_2009, witten_properties_1984, conlon_qcd_2006, arvanitaki_string_2010, cicoli_type_2012, workman_review_2022, irastorza_introduction_2022, irastorza_new_2018, abramowski_constraints_2013, masso_axions_2008}. \Acp{alp} do not necessarily solve the strong \ac{cp} problem, but bear many phenomenological similarities to axions, making them also viable \ac{dm} candidates as \acp{wisp}. 
\ifthenelse{\boolean{isthesis}}
    {A comprehensive introduction to the physics of axions and \acp{alp} can be found in \cref{app:alp_bkg}.}
    {}

As will be seen in \cref{sec:alp_interact}, the coupling between photon pairs and \acp{alp} means that \acp{alp} can interconvert with photons in the presence of an external magnetic field. This would lead to \alpphot mixing, a process which is mathematically closely analogous to neutrino oscillations \cite[e.g.,][]{chadha-day_axion-like_2022}.
This mixing could distort the photon spectra from astrophysical sources \cite{hooper_detecting_2007}. As an example, photons could be converted into \acp{alp} in a magnetic field at or near a source, or in some intervening field encountered on the trajectory from the source to the observer, such as the \ac{igmf}. The \alpphot beam could then undergo further oscillations, including from \ac{alp} back to photon, in the Galactic magnetic field. This \alpphot mixing may lead to irregularities in the observed spectrum of a source \citep[see, e.g.,][for a comprehensive review]{biteau_gamma-ray_2022}. 
%
%\note{\Colin{maybe it would make sense to have this part come first, starting with the enhanced transparency, and then ending with the oscillatory feature case}}

One such irregularity would be an increased probability that high-energy \graysnoh{} survive the pair absorption process with the \ac{ebl}. This would effectively increase the transparency of the Universe to \graysnoh{}, leading to unusually hard \gray{} spectra and potentially the detection of \gray{} sources out to greater distances. 
Another spectral irregularity would manifest as oscillatory features in the observed spectrum\textemdash a result of the energy dependence of the \alpphot mixing. 
%The coupling between photons and \acp{alp} becomes stronger with increasing magnetic field. 
The \acs*{alp}-induced spectral irregularity becomes more prominent when the photons travel through a stronger magnetic field or across a larger magnetic field domain. 
Therefore, stronger experimental constraints on the properties of \acp{alp} can be obtained from observations of \gray sources that are either %distant or embedded in a strong external magnetic field. \Colin{[Maybe rewrite to: "obtained from observations of \gray sources that are either 
embedded in a relatively strong magnetic field, or distant enough that their emission would extensively traverse through the \ac{igmf}. %" or something like that]}

In addition to photons, axions would also couple with gluons and, in some models \citep[e.g., the \acs{dfsz} model;][]{zhitnitsky_possible_1980, dine_simple_1981}, with fermions, such as electrons. In hadronic models \citep[e.g., the \acs{ksvz} model;][]{kim_weak-interaction_1979, shifman_can_1980}, the coupling between axions and electrons is negligible. \Acp{alp}, however, may generally couple to electrons. In this work, we assume that the coupling between \acp{alp} and both gluons and electrons is subdominant.

%\info[inline]{\Qi{(placeholder) maybe this is a good place to introduce NGC~1275, cool-core (I think they ditched this name, it's called bright core now??) clusters, and also discuss \ac{igmf}. }}

%\Qi{(work in progress) \ac{igmf} is very weak, below $\mathcal{O}{(1\text{nG})}$ with various coherence lengths assumed \citep{POLARBEAR2015IGMF, Neronov21IGMF, Padmanabhan23IGMF_FRB}. Bright-core galaxy clusters, on the other hand, are shown by radio rotation measure to have much stronger magnetic field than \ac{igmf}. NGC~1275, located at the center of the Perseus Cluster, may be surrounded by magnetic field even stronger?? (need to look up if this is true or not)}

Several experiments have directly searched for \acp{alp} using various methods \citep[see, e.g.,][for a review]{irastorza_new_2018}. Purely laboratory experiments, the so-called \ac{lsw} experiments such as \ac{alps} I \cite{ehret_new_2010}, search for photons that appear to pass through a wall by oscillating into intermediate \acp{wisp} and have established upper limits on the \alpphot coupling strength at the level of $\axg< \mathcal{O}{(10^{-8})}\; \mathrm{GeV}^{-1}$ at masses $m_a\lesssim1$ meV. The \ac{alps} II experiment \cite{bahre_any_2013} is expected to improve the limits to $\axg\lesssim 2\times10^{-11}\; \mathrm{GeV}^{-1}$ in a similar mass range. Helioscopes and haloscopes seek to convert \acp{alp} from the Sun and the Galactic \ac{dm} halo, respectively, into detectable photons to constrain the coupling strength between \acp{alp} and photons. The \ac{cast}, a helioscope, has looked for solar axions converting into X-rays in a laboratory magnetic field and established an upper limit on the axion-photon\footnote{By virtue of the shared phenomenology between axions and \acp{alp}, all the constraints discussed in this section may be generalized to \acp{alp}.} coupling strength of $\axg\lesssim 6.6\times10^{-11}\; \mathrm{GeV}^{-1}$ at mass $m_a\lesssim0.02\;$eV \cite{cast_collaboration_new_2017}. 
The \ac{admx}, a haloscope, has searched for dark matter axions converting into microwave photons in a large resonant cavity within a strong superconducting magnet and established upper limits on the axion-photon coupling strength of $\axg < \mathcal{O}{(10^{-13})}\; \mathrm{GeV}^{-1}$ within a narrow mass range of $2.66\lesssim m_a\lesssim2.80 \;\mu\mathrm{eV}$ \cite{admx_collaboration_search_2018}. To constrain the coupling between \acp{alp} and gluons and nuclear spin, the \ac{casper} uses nuclear magnetic resonance to detect an oscillating torque on nuclear spins \cite{budker_proposal_2014}. Constraints on \acp{alp} have also been derived from observations of other astrophysical sources, such as giant stars \cite{xiao_constraints_2021}, neutron stars \cite{berenji_constraints_2016}, including magnetars \cite{fortin_magnetars_2021}, the supernova SN~1987A \citep[e.g.,][]{payez_revisiting_2015, lee_revisiting_2018}, globular clusters \cite{lucente_constraining_2022}, spinning black holes \cite{chen_stringent_2022}, and \acp{agn}, as well as from various cosmological observations \citep[e.g.,][]{marsh_axion_2016}. 

Constraints on \acp{alp} using spectral irregularities from \acp{agn}, including blazars and \aclp*{rg}, have been derived from X-ray and \gray{} observations. These constraints improved the upper limit on the \alpphot coupling strength $\axg$ by more than an order of magnitude compared to the limits from \ac{cast}, covering an \ac{alp} mass range ($\lesssim\mu$eV) lower than what \ac{admx} is sensitive to \citep[e.g.,][]{ohare_cajohareaxionlimits_2020}. 
Blazars and \acp{rg} exhibit non-thermal radiation powered by the central supermassive black hole, featuring smooth X-ray and \gray{} spectra that can often be described by a power law or log parabola model \citep[e.g.,][]{bhatta_hard_2018, ajello_fourth_2022}. Blazars and \acp{rg} that reside in the center of galaxy clusters are particularly interesting for \ac{alp} studies, as the \ac{icm} is permeated by large-scale turbulent magnetic fields where \alpphot conversion may take place efficiently. These fields are particularly strong close to the center in cool-core clusters \cite{cicoli_355_2014}, where the \ac{icm} is so dense that the cooling time, primarily from free-free Bremsstrahlung \cite{lea_thermal-bremsstrahlung_1973}, is short compared to the Hubble time \cite{cowie_radiative_1977, fabian_subsonic_1977}, resulting in a rapidly cooled \ac{icm} \cite{hudson_what_2010, lehle_heart_2024}.

The TeV \acl*{rg} NGC~1275, located at the center of the cool-core Perseus Cluster at a distance of ${\sim 75~{\rm Mpc}}$ (${z = 0.01756}$), has been extensively studied to search for evidence for \acp{alp}. The Perseus Cluster is the brightest galaxy cluster in the X-ray band, with a strong \ac{icmf} of $\mathcal{O}{(10{\rm~\mu G})}$, as shown by radio \acp{rm} \cite{taylor_magnetic_2006}. In comparison, the \ac{igmf} is much weaker, constrained by several methods to be below $\mathcal{O}{(1{\rm~nG})}$ with various coherence lengths assumed \cite{kronberg_extragalactic_1994, blasi_cosmological_1999, polarbear_collaboration_polarbear_2015, neronov_limit_2023, padmanabhan_new_2023}. 
The Chandra X-ray Observatory and the \Fermi-\ac{lat} have searched for spectral irregularities in NGC~1275's X-ray and GeV \gray{} emission, respectively, to set limits on the \alpphot coupling strength ($\axg \lesssim 6\times10^{-13}\; \mathrm{GeV}^{-1}$ at masses $m_a\lesssim1\times10^{-12}$ eV from X-ray, and $\axg \lesssim 5\times10^{-12}\; \mathrm{GeV}^{-1}$ in the mass range of $0.5\;\mathrm{neV}\lesssim m_a\lesssim5 \;\mathrm{neV}$ from \gray{}) \cite{reynolds_astrophysical_2020, sisk-reynes_new_2021, ajello_search_2016}. 

At TeV \gray{} energies, the exceptional sensitivity on short timescales from ground-based \acp{iactel} offers an opportunity to search for spectral irregularities in fast and intense flares from blazars and \aclp*{rg}. 
% The \acf{hess} has used TeV \gray{} observations of the blazar PKS~2155-304, located at the center of a galaxy cluster, to constrain $\axg \lesssim 2.1\times10^{-11}\; \mathrm{GeV}^{-1}$ in the mass range of $15\;\mathrm{neV}\lesssim m_a\lesssim60 \;\mathrm{neV}$ \cite{abramowski_constraints_2013}. Unlike the searches performed with Chandra and \Fermi-\acs{lat} data, which used the data over several years, only 15 hours of \ac{hess} data over five days were chosen, focusing on an intense flaring state to maximize the \acl*{s/n}.
\hiddentodo[inline]{Compare the limits obtained from spectral irregularities and laboratory experiments? (great suggestion, I added at the beginning of the paragraph above)}
\hiddentodo[inline]{add more reasoning for choosing NGC~1275 and why this paper adds to previous results? }
An exceptionally bright \ac{vhe} flare of NGC~1275 was recorded by the \acf{magic} beginning on the night spanning December 31\textsuperscript{st}, 2016 to January 1\textsuperscript{st}, 2017 (MJD~57753-57754) \cite{mirzoyan_magic_2017}. The brightest outburst, recorded during the first night of the flare, had a flux $\sim 1.5$ times the flux of the Crab Nebula above 100 GeV, falling on subsequent nights \cite{ansoldi_gamma-ray_2018}, during which it was further detected by the \acf{vts} \cite{mukherjee_veritas_2017}. 
Observations of this flare present an excellent opportunity to search for \acp{alp} through \gray{} spectral irregularity, as (1) the source is embedded in a strong \ac{icmf} and (2) the bright flare allows for a well-characterized TeV \gray{} spectrum due to the superior \acl{s/n}. Motivated by this, \ac{magic} used their flaring and post-flare data from NGC~1275, alongside long-term quiescent monitoring data of the source, to exclude at 99\% confidence \ac{alp} coupling constants ${\axg \gtrsim 3 \times 10^{-12} {\rm~GeV^{-1}}}$ in the range ${4 \times 10^{-8} {\rm~eV} \lesssim \axm \lesssim 9 \times 10^{-8} {\rm~eV}}$ \cite{abe_constraints_2024}.

During \ac{vts} observations of NGC~1275 spanning January 2\textsuperscript{nd} (MJD~57755) to January 3\textsuperscript{rd} (MJD~57756) of 2017, an average flux state at 50\% of the Crab flux at TeV energies was recorded. During the flare, NGC~1275 was detected by \ac{vts} over 2.2 hours (MJD~57755) with three-telescope operations and 1.3 hours (MJD~57756) with four-telescope operations, noting a marginally higher measured flux on the first night observed. The two nights recorded respective statistical significances of $\sim 31 \sigma$ and $\sim 22 \sigma$. These observations overlap with a Bayesian-block-identified steady state in \Fermi-\acs{lat} spanning roughly January 1\textsuperscript{st} to January 6\textsuperscript{th} of 2017 \cite{rulten_in_prep}. However, as the \ac{vhe} spectra recorded by \ac{magic} appear to evolve significantly over this period \cite{ansoldi_gamma-ray_2018, abe_constraints_2024}, we opt to exclude the Bayesian-block \Fermi-\acs*{lat} data from this study. This disconnect between the evolution of the flare in GeV and TeV bands may be explained by stronger cooling at \acp{vhe}. 
%Provided the better \ac{vhe} statistics associated with the higher flux on January 2\textsuperscript{nd}, we opt to include only the \acs{vts} data from January 2\textsuperscript{nd} in this analysis.
In addition, to minimize evolution in the \ac{vhe} band, we opt to include only \acs{vts} data from January 2\textsuperscript{nd} in this analysis due to the higher VERITAS observed flux on this date.
\hiddentodo[inline]{Add: maybe something about the dates following what we hear from the NGC~1275 paper people} %between December 31, 2016
%\unsure{I think the flare started Dec 31, but VTS only got on it January 2 (also note that I only use the Jan 2 data for the analysis)} and January 3, 2017.

In this work, we use these \acs{vts} observations to set limits on the \alpphot coupling strength and the mass of \acp{alp}.
%, despite great interest in the source with the upcoming Cherenkov Telescope Array Observatory (CTAO) \cite{abdalla_sensitivity_2021}. 
%The limits we derive are more constraining than previously reported limits using \Fermi-\acs{lat} observations of blazars and NGC~1275.\change{JH: Bit arrogant, convey this in a kinder way.} %which excluded $\axg\gtrsim3\times10^{-12}\;\mathrm{GeV}^{-1}$ at the \ac{alp} mass $m_a\sim10^{-7}$ eV with a confidence level of 99\%.\unsure{isn't this basically the same as our exclusion region? I feel like the previous limits were at slightly lower masses.. which papers are being referenced here?}}
\hiddentodo[inline]{discussion of flare, why flare was so dramatic, why flares are useful for this study}
In \cref{sec:alp_interact} we discuss the theory of \alpphot interactions. Then, in \cref{sec:dataana}, we describe the \ac{vts} observatory that recorded the data and the techniques used to analyze them. In \cref{sec:statmethods}, we delineate the statistical methods employed to compare the \ac{alp} and no-\acs*{alp} hypotheses.
In \cref{sec:alp-results}, we report our findings, and compare our methodology and results to those reported previously by \ac{magic} \citep{abe_constraints_2024}.

\section{ALP-Photon Interactions}
\label{sec:alp_interact}
\hiddentodo[inline]{\citet{galanti_axion-like_2022} does the best job of going through the \ac{alp} theory case by case; simplified all the way to the general case. So does \citet{tavecchio_evidence_2012} and \citet{gallardo_romero_search_2020} and \citet{galanti_extragalactic_2018}}
\hiddentodo[inline]{\citet{davies_relevance_2021} does the best job of going through the contributions to the $\Delta_{\gamma\gamma}$}
\hiddentodo[inline]{\citet{galanti_blazar_2019} has some important notes about $\mathcal{L}_{\rm HEW}$ and some Dobrynina reference. Also the limitations of photon dispersion on the CMB et al.}

The theory of \alpphot interactions has been covered extensively in previous publications \citep[see, e.g.,][among many others]{raffelt_mixing_1988, bassan_axion-like_2010, de_angelis_relevance_2011, meyer_detecting_2014, galanti_extragalactic_2018, galanti_axion-like_2022, gallardo_romero_search_2020, davies_relevance_2021, matthews_how_2022, kachelries_origin_2022}. An abbreviated summary is provided here\ifthenelse{\boolean{isthesis}}
  {, and is further expanded on in \cref{app:alp_phot_interact}.}
  {.}

The \ac{alp} contribution to the \ac{sm} Lagrangian\footnote{Using natural Lorentz-Heaviside units such that $\hbar = c = k_{\rm B} = 1$ throughout, and $\alpha = e^2/4\pi \simeq 1/137$.} is given by \cite{raffelt_mixing_1988, bassan_axion-like_2010, kachelries_origin_2022}:
\begin{equation}\label{eqn:lagr}
    \mathcal{L}_{\rm ALP} = \underbrace{\frac{1}{2} \partial_{\mu} a \partial^{\mu} a - \frac{1}{2}\axm^{2}a^{2}}_{\mathcal{L}_{aa}} + \underbrace{\frac{1}{4}\axg F_{\mu \nu}\Tilde{F}^{\mu \nu}a}_{\mathcal{L}_{a \gamma}},
\end{equation}
%where $a$ is the ALP field, 
%, $\axm$ and $\axg$ are its mass and coupling strength to photons, respectively, 
${F_{\mu \nu}}$ is the \ac{em} field tensor, and ${\Tilde{F}^{\mu \nu} = \frac{1}{2}\varepsilon_{\mu\nu\rho\sigma}F^{\rho\sigma}}$ is its dual. 
The contribution $\mathcal{L}_{aa}$ contains a kinetic and mass term to describe \acp{alp} as a non-interacting, free scalar field $a$ with mass \axm \cite{bassan_axion-like_2010, arias-aragon_axion-like_2023}.
The rightmost term describes the \ac{alp} coupling to electromagnetism through the interaction,
\begin{equation}\label{eqn:Linter}
    \mathcal{L}_{a \gamma} = \frac{1}{4}\axg F_{\mu \nu}\Tilde{F}^{\mu \nu}a = 
    \axg \left(\mathbf{E} \cdot \mathbf{B}\right) a,
\end{equation}
where $\axg$ is the coupling strength of the \ac{alp} field to photons. As a consequence, any photon in the presence of a magnetic field with some component oriented transversely to the photon propagation direction will mix with \acp{alp} \cite{raffelt_mixing_1988}.

Because of the coupling described in \cref{eqn:Linter}, an \alpphot beam will propagate as a mixing of three quantum states: two corresponding to photon polarization, and the third corresponding to the \ac{alp} \cite{abramowski_constraints_2013}. For an initially polarized photon beam of energy $E$ propagating through a single homogeneous magnetic field $B$ (with component $B_T$ that is transverse to photon propagation direction, and parallel to the polarization state), the equations of motion for an \alpphot beam can be solved \citep[see, e.g.,][]{raffelt_mixing_1988, de_angelis_relevance_2011, galanti_behavior_2018}, and result in \alpphot oscillations with amplitude:

\begin{equation}\label{eqn:oscprob}
    P_{\gamma \rightarrow a}(E,z) = \left( \frac{\axg B_T}{\Delta_{\rm osc}(E)} \right)^2 \sin^2 \left( \frac{\Delta_{\rm osc}(E)(z-z_0)}{2} \right),
\end{equation}
for a beam that has traveled from $z_0$ to $z$. The oscillation wave number is given by \cite{davies_constraints_2023}

\begin{equation}\label{eqn:deltaosc}
\ifthenelse{\boolean{isthesis}}
{
    \Delta_{\rm osc}(E) = \left[ \left\{ \frac{m_a^2 - \omega_{\rm pl}^2}{2 E} + E \left( \frac{7}{2} b + \chi + i \frac{\Gamma_{\gamma \gamma}}{2 E} \right) \right\}^2 + (\axg B_T)^2 \right]^{1/2},
}
{
    \begin{aligned}
    &\Delta_{\rm osc}(E) 
    \\
    &= \left[ \left\{ \frac{m_a^2 - \omega_{\rm pl}^2}{2 E} + E \left( \frac{7}{2} b + \chi + i \frac{\Gamma_{\gamma \gamma}}{2 E} \right) \right\}^2 + (\axg B_T)^2 \right]^{1/2},
    \end{aligned}
}
\end{equation}
where ${\omega_{\rm pl}}$ is the plasma frequency ${\omega_{\rm pl} \sim 0.037 \sqrt{n_e}~{\rm neV}}$, with the electron density $n_e$ in units of ${\rm cm}^{-3}$. $\chi$ and $\Gamma_{\gamma \gamma}$ give the terms for the dispersion and absorption for the surrounding photon fields, respectively, and $b$ is the vacuum \ac{qed} term which describes dispersion of the magnetic field:

\begin{equation}\label{eqn:bterm}
    b = \frac{\alpha}{45 \pi} \left ( \frac{B_T}{B_{\rm cr}} \right )^2 ,
\end{equation}
\begin{sloppypar}
\noindent where $\alpha$ is the fine structure constant, and $B_{\rm cr}$ is the critical magnetic field strength ${B_{\rm cr} = m_e^2/\abs{e} \approx 4.41 \times 10^{13}~\mathrm{G}}$ with $m_e$ the electron mass and $e$ the fundamental electric charge.
\end{sloppypar}

Assuming that the absorption is small, it is possible to define a critical energy \cite{de_angelis_relevance_2011, gallardo_romero_search_2020, davies_constraints_2023}
\begin{equation}\label{eqn:Ecrit}
    E_{\rm cr}^{\rm low} = \frac{\abs{m_a^2 - \omega_{\rm pl}^2}}{2 \axg B_T},
\end{equation}
far above which the conversion probability is maximal and independent of energy (the \textit{strong-mixing regime}), and below which the conversion probability is oscillatory, and increasing, with energy. The strong-mixing regime is bounded from above by

\begin{equation}\label{eqn:Ehigh}
    E_{\rm cr}^{\rm high} = \frac{\axg B_T}{\frac{7}{2}b + \chi}.
\end{equation}
%, as noted before, contributions arising from \ac{qed} vacuum polarization, the CMB, and absorption are negligible for the astrophysical environment under consideration in this paper \cite{de_angelis_relevance_2011}. 
The energy-dependent oscillatory behavior near $E_{\rm cr}^{\rm low}$ is of particular interest for this study, as these oscillations have the potential to imprint features in the spectra of astrophysical \gray{} emitters. 

\subsection{Modeling the ALP Effect}\label{sec:alpmodel}
The above is a simple and illustrative example of the effects one can expect from \alpphot mixing; however, for the purposes of this paper it is necessary to consider the more complex scenario of an initially unpolarized \alpphot beam traversing non-homogeneous magnetic fields. This can be properly accounted for by splitting magnetic regions into slices of homogeneous field and by using the formalism of the density matrix \cite{de_angelis_relevance_2011, mirizzi_stochastic_2009,meyer_detecting_2014, kartavtsev_extragalactic_2017, ajello_search_2016}. The assumption of an initially unpolarized beam is necessary because, although the beam will likely have some degree of polarization, it is not known \textit{a priori}, nor do current-generation \gray{} telescopes have the ability to measure it \cite{galanti_comment_2013}. Further, it is expected to be small\textemdash $\mathcal{O}({\rm a~few~} \%)$ \cite{tavecchio_privcomm_2023}.

We model the propagation of the \alpphot beam as it traverses three distinct environments. Since NGC~1275 is located at the center of the Perseus Galaxy cluster, the first region the beam will encounter will be the \ac{icmf} (see \cref{sec:perseus} for details). 
Provided the relatively small domain traversed by the emission (considering that NGC~1275 is a \acl*{rg} observed at large viewing angles with respect to the jet axis \cite{fujita_discovery_2017, abdalla_sensitivity_2021}) and the relative dominance of the cluster contribution to mixing over the jet component \cite{davies_relevance_2021}, we are able to neglect \alpphot mixing in the \ac{agn} jet. We next include \ac{ebl} absorption for propagation through intergalactic space, using the model of \textcite{dominguez_extragalactic_2011}. 
As noted in \cref{sec:alpintro}, current limits place the maximum strength of the \ac{igmf} at $\sim 10^{-9}~{\rm G}$. Given the additional constraint $\axg < 6.6 \gunit$ \cite{cast_collaboration_new_2017}, we find that ${E_{\rm cr}^{\rm low} \lesssim 3.5~{\rm TeV}}$ only for ${\axm \lesssim 3 \times 10^{-9}~{\rm eV}}$. This is below the \axm range considered for this analysis, so we only consider the regime of \axm where the ${E_{\rm cr}^{\rm low}} \gtrsim 3.5~{\rm TeV}$, which is roughly the upper bound of our instrument's sensitivity in this analysis. Together with the relatively small redshift of NGC~1275, we do not expect strong irregularities in the spectrum to be induced by mixing in the \ac{igmf}, and therefore neglect its contribution as well.
Finally, we consider beam propagation and mixing in the \acl*{mw}'s magnetic field, following the model of \textcite{jansson_new_2012}.

In order to compute the probability to observe at Earth a $\gamma~{\rm ray}$ (of either polarization) from an initially unpolarized beam at the source, known as the photon survival probability $P_{\gamma \gamma}$, we use the \pkg{gammaALPs}\footnote{\url{https://github.com/me-manu/gammaALPs}} software package \cite{meyer_manuel_gammaalps_2021}. The code numerically solves the equations of motion for the \alpphot system using the transfer-matrix formalism and incorporates all relevant mixing terms in the mixing matrix, including dispersion from \ac{qed} effects and the \ac{cmb}. 

\subsubsection{Perseus Cluster Modeling}\label{sec:perseus}
Following from Refs.\@ \cite{meyer_detecting_2014, ajello_search_2016}, we expect the \ac{icmf} of the Perseus cluster to be a turbulent field, with a strength that traces the electron density $n_{e}(r)$ of the \ac{icm}, ${B(r) = B_{0}[n_{e}(r)/n_{e}(r=0)]^{\eta}}$ \cite{dolag_correlation_2001}. The turbulent component is modelled as a divergence-free homogeneous isotropic field with Gaussian turbulence with a mean of zero and variance $\sigma_B$. 
We assume that the power spectrum of the turbulence follows a power law ${M(k) \propto k^{q} }$ in wave numbers $k$ between the maximum and minimum turbulence scales ${k_{L} = 2\pi/\Lambda_{\rm max}}$ and ${k_{H} = 2\pi/\Lambda_{\rm min}}$, and 0 otherwise \cite{meyer_detecting_2014, kachelries_origin_2022}.
We use an analytical approximation for $n_{e}(r)$ derived from X-ray observations with \textit{XMM-Newton} (Eq. (4) in \textcite{churazov_xmmnewton_2003}) within 500 kpc of the cluster's center ($r_{\rm max}$), and conservatively assume zero magnetic field beyond\footnote{For \ac{alp} parameters and photon energies relevant to \acp{iactel}, a typical \alpphot oscillation length for the Perseus Cluster is ${\mathcal{O}(10~{\rm kpc})}$ \cite{abramowski_constraints_2013}.}. \Acp{rm} determined from \ac{vlba} observations of the innermost region (tens of \acs{pc}) around NGC~1275 suggest a central magnetic field for the cluster of $25~{\rm \mu G}$ \cite{taylor_magnetic_2006}. Considering this, alongside an independently derived lower limit on ${B_{0} \gtrsim 2-13~{\rm \mu G}}$ for ${0.3 \leq \eta \leq 0.7}$ from the \acs{magic} observatory \cite{aleksic_constraining_2012}, we make the assumptions ${\sigma_{B}=10~{\rm \mu G}}$ and ${\eta = 0.5}$, which are consistent with measurements of similar cool-core clusters Hydra A \cite{kuchar_magnetic_2011} and A2199 \cite{vacca_intracluster_2012}. In the absence of detailed \acp{rm} outside of the central-most region of the Perseus cluster, we assume values from A2199, which has a comparable number of member galaxies. A summary of the fiducial parameters used to model the magnetic field structure of the Perseus cluster can be found in Table I of \textcite{ajello_search_2016}.

\hiddentodo[inline]{need a short version of this maybe look at \citet{abramowski_constraints_2013}. \citet{li_limits_2021} does it pretty quickly. Maybe also \citet{hooper_detecting_2007}. \citet{li_searching_2021} also gets through it pretty quickly. \citet{cheng_revisiting_2021} also pretty quick. maybe also \citet{bassan_axion-like_2010}. \citet{horns_hardening_2012} also pretty short.}
% evolve according to a second-order wave equation,
% a photon of energy $E$ propagating in a homogeneous field $B$ (with a component $B_T$ that is transverse to the direction of photon propagation, and parallel to one of its polarization states) will 

\section{Data Analysis}
\label{sec:dataana}
Data for this study were recorded by \ac{vts}, a ground-based \gray observatory comprising an array of four \acp{iactel} located in southern Arizona at 31$^\circ$40'30"N 110$^\circ$57'07"W, 1.3 km \acs{asl} and capable of detecting \graysnoh{} from 85 GeV to $>$30 TeV \cite{park_performance_2016}.
%An energy resolution of $\sim$15\% (at 1~TeV) and a point spread function of $\sim$0.1$^{\circ}$ (68\% containment, at 1~TeV) can be achieved with \acs{vts} observations. \acs{vts} is able to detect a point source of 1\% the Crab Nebula flux at a statistical significance of five standard deviations (5 $\sigma$) in $\sim$25~hours.}\change{probably can condense this section a little}
\acs{vts}, similar to other current ground-based \gray{} observatories, offers moderate energy resolution ($\sim$15\% at 1~TeV) limited by the air shower fluctuations associated with the air Cherenkov technique \citep[e.g.,][]{hofmann_improved_2000}.

For this study, we used \acs{vts} observations of NGC~1275 in a flaring state on January 2\textsuperscript{nd}, 2017 with an exposure of 2.2 hr after data quality selection.
%Fermi-LAT Bayesian blocks: 
%BB1 504834750 - 504899550 = 2016-12-30 23:52:26.000 UTC - 2016-12-31 17:52:26.000 UTC
%BB2 504899550 - 505396350 = 2016-12-31 17:52:26.000 UTC - 2017-01-06 11:52:25.000 UTC
% BB1 Fermi is used together with Jan 1 MAGIC
% BB2 Fermi is used together with Jan 2 \acs{vts}
Spectral analysis for these data was performed using the \pkg{gammapy} software package \cite{acero_fabio_gammapy_2023}. \acs{vts} data were pre-processed using the \acl{ed} \acs{vts} analysis pipeline \cite{maier_eventdisplay_2017} with a \acl{rr} background model \cite{berge_background_2007} and were provided to \pkg{gammapy} at the event level, a \acl{dl}~3 (\acs{dl}3) product \citep[see][]{nigro_evolution_2021}. From the \acs{dl}3 data, a full region-based on-off spectral analysis was performed.

Two different intrinsic models for the data were examined for the null hypothesis case, an \ac{ecpl} and a \ac{lp}, neither of which was significantly preferred over the other. %\textcite{rulten_in_prep}\unsure{Paper is not going to be Rulten, but a VERITAS paper}
%In an ongoing study \unsure{Paper is not going to be Rulten, but a VERITAS paper} using the same \ac{vts} data with \Fermi-\ac{lat} data that overlaps but is not strictly simultaneous, 
In an ongoing study using the same VERITAS data that overlaps with the integrated \Fermi-\ac{lat} data for January 2\textsuperscript{nd} \cite{rulten_in_prep}, 
a \ac{bic} was utilized to find that a \ac{lp} fit was preferred to an exponentially cutoff power law. Therefore, we have chosen the \ac{lp} as the base intrinsic model for our analysis:
\begin{equation}\label{eqn:lp}
    \Phi_{\rm int}(E) = N_0 \left( \frac{E}{E_0} \right)^{-\alpha -\beta \log (E / E_0)},
\end{equation}
where $\Phi_{\rm int}(E)$ is the intrinsic differential photon flux 
%number of photons received per unit area per unit time 
at photon energy $E$, $N_0$ the spectral normalization, $E_0$ the reference energy, $\alpha$ the spectral index, and $\beta$ the curvature. The null hypothesis model is then given by
\begin{equation}\label{eqn:nullhyp}
    \Phi_{0}(E,z) = e^{-\tau(E,z)}\Phi_{\rm int}(E),
\end{equation}
where $\tau(E,z)$ is the optical depth due to the \ac{ebl}, from \textcite{dominguez_extragalactic_2011}. For the alternative hypothesis, the photon survival probability ${P_{\gamma \gamma}(E, z; \axm, \axg, \bm{B})}$, which includes both \ac{ebl} attenuation and the \ac{alp} effect (dictated by \axm, \axg, and the magnetic field environment traversed by the beam $\bm{B}$), will be used instead, yielding  
\begin{equation}\label{eqn:alphyp}
    \Phi_{\rm ALP}(E, z; \axm, \axg, \bm{B}) = P_{\gamma \gamma}(E, z; \axm, \axg, \bm{B})\Phi_{\rm int}(E).
\end{equation}
${P_{\gamma \gamma}(E, z; \axm, \axg, \bm{B})}$ is calculated with the \pkg{gammaALPs} \cite{meyer_manuel_gammaalps_2021} software package.

The models given by \cref{eqn:nullhyp,eqn:alphyp} can be used to compute their respective likelihood given the observed data. The likelihood, across all energy bins $i$, can be written as
\begin{equation}\label{eqn:poiss_lik}
    \Lik (\mu (\axm, \axg, \bm{B}, \bm{\theta} ), b | \bm{D} ) = \prod_i \Lik_i ( \mu_i (\axm, \axg, \bm{B}, \bm{\theta} ), b_i | \bm{D_i} ),
\end{equation}
where $\mu_i$ is the expected number of counts in the $i$-th energy bin, computed by forward folding the detector response into the model given by \cref{eqn:nullhyp} or \labelcref{eqn:alphyp}. This number depends on \axm, \axg, the realization of the magnetic field $\bm{B}$, and the \ac{sed} nuisance parameters which define \cref{eqn:lp}: $\bm{\theta} = (N_0, \alpha, \beta)$. The parameter $b_i$ is the number of background counts expected in the source region, and $\bm{D_i}$ the counts measured by the instrument in its source and background estimation regions, also for the $i$-th energy bin.

As suggested above, there are two cases under consideration: the no-\acs*{alp} case (the null hypothesis), corresponding to the case where \acp{alp} do not couple to photons ($\axg = 0$), and the \ac{alp} case (the alternative hypothesis). Following from \cref{eqn:poiss_lik}, the likelihood of the alternative hypothesis can be written as
\begin{equation}
    \Lik_{\rm ALP} (\mu (\axm, \axg, \bm{B}, \bm{\theta} ), b | \bm{D} ),
\end{equation}
and the likelihood of the null hypothesis can be written simply as
\begin{equation}
    \Lik_0 (\mu (\bm{\theta} ), b | \bm{D} ).
\end{equation}

For the fixed parameters (\axm, \axg), the nuisance parameters can be ``profiled out'' using the profile likelihood method of \textcite{rolke_limits_2005}. This profile approach is employed by the WStat fit statistic used in \pkg{gammapy} \cite{wachter_parameter_1979}. Profiling is done for the expected background counts $b_i$ by replacing them with the value $\hat{b}_i$ that analytically maximizes the likelihood, where $\hat{b}_i$ is effectively a background estimator. This technique is known to underestimate the predicted background counts, particularly in the higher-energy regime for \acp{iactel}, where the number of background counts will likely drop to zero \cite{buchner_statistical_2022}. This has been addressed by adaptively rebinning the data, ensuring that each energy bin contains $> 0$ background counts. Although this does not eliminate bias in the predicted counts, the relative scale of the bias is significantly reduced and, when tested on fits of simulated null-hypothesis data, was shown to adequately recover the injected model. The \ac{sed} nuisance parameters $\bm{\theta}$ are profiled for a fixed (\axm, \axg) by performing an \pkg{iMinuit} \citep{dembinski_scikit-hepiminuit_2024, james_minuit_1975} minimization\footnote{The WStat fit statistic implemented in \pkg{gammapy} follows $-2 \ln \Lik$.} of \pkg{gammapy}'s WStat fit statistic with the expected background counts profiled, yielding $\bm{\hat{\theta}}$.

% \note{Note: it's \url{https://github.com/gammapy/gammapy/blob/ecfdff6d5a7042e649c4651a28aa1264adebe8bd/gammapy/datasets/flux\_points.py\#L291} for \Fermi-\acs{lat} and \url{https://github.com/gammapy/gammapy/blob/ecfdff6d5a7042e649c4651a28aa1264adebe8bd/gammapy/datasets/map.py\#L2160-L2165} for \acs{vts} summed up over bins} 

To treat the magnetic field structure of Perseus statistically, for each (\axm,~\axg) we perform fits over $N_B=100$ random realizations of the turbulent magnetic field $\bm{B}_j$, $j=1,\ldots,N_B$, where each individual fit yields, following the profiling described above, $\Lik_{\rm ALP}(\mu (\axm, \axg, \bm{B_j}, \bm{\Hat{\theta}} ), \Hat{b} | \bm{D})$. However, it is highly unlikely that the magnetic field that is realized in nature is included in the simulated magnetic-field realizations. Therefore, instead of profiling over the magnetic-field realizations, we sort them by their $\Lik_{\rm ALP}$, choosing the realization $j$ associated with the 95\% most likely fit, following from Refs.\@ \cite{abdalla_sensitivity_2021,ajello_search_2016, davies_constraints_2023}. For $N_B=100$ realizations, this corresponds to $j=95$, or the quantile $Q_B = 0.95$\footnote{Note that this is the 0.05 quantile of $-2 \ln\Lik_{\rm ALP}$.}. The selected magnetic field realization can be denoted $\bm{B_{95}}$, taking particular care to note that $\bm{B_{95}}$ may be different for different \ac{alp} parameters, such that $\bm{B_{95}}\equiv\bm{B_{95}}(\axm,~\axg)$. The maximum likelihood for a fixed (\axm,~\axg) will then be given by ${\Lik_{\rm ALP}(\mu (\axm, \axg, \bm{B_{95}}, \bm{\hat{\theta}} ), \hat{b} | \bm{D})}$.

\begin{figure}
    \centering
    \includegraphics[width=\columnwidth]{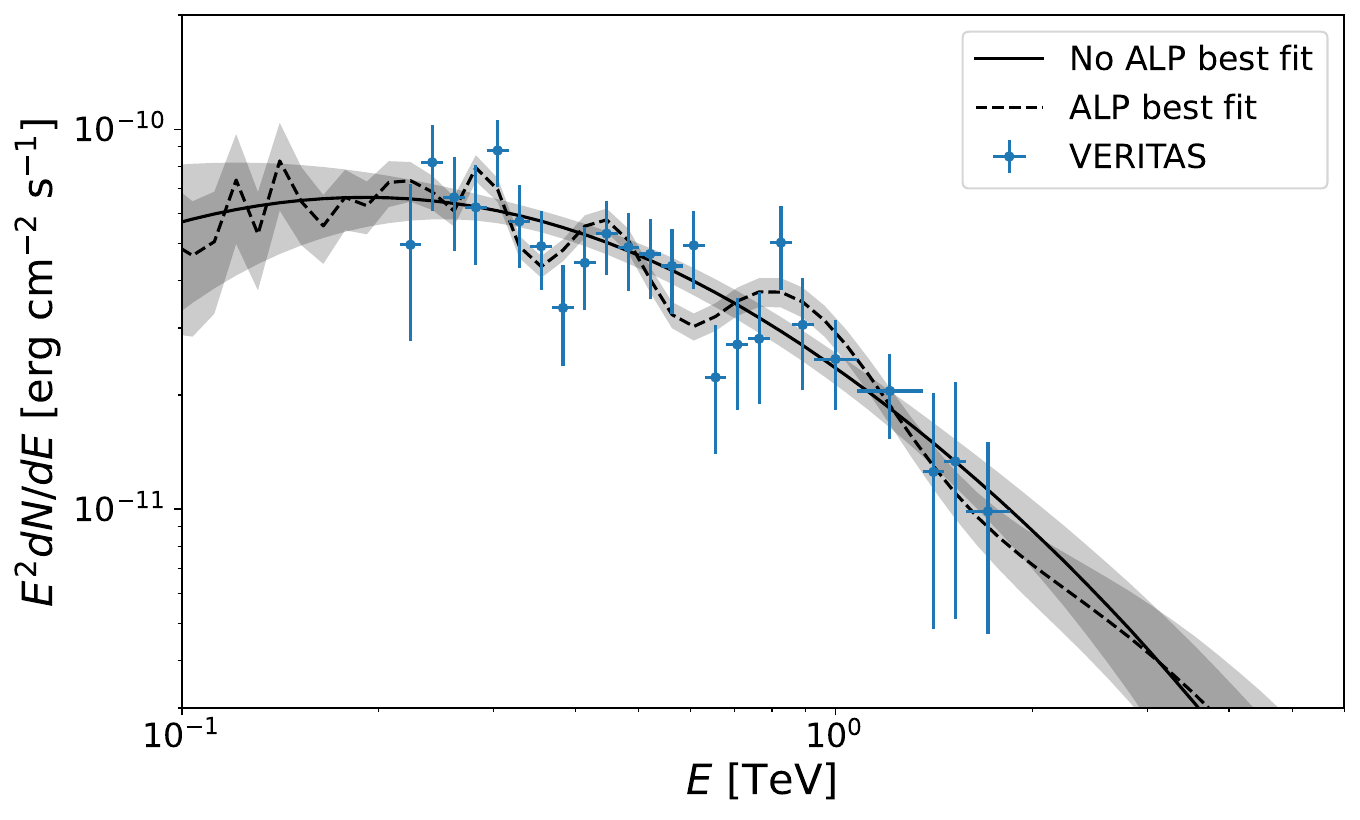}
    \caption[\Acl*{sed} for the January 2\textsuperscript{nd}, 2017 observations of the NGC~1275 flare.]{\Ac{sed} for the January 2\textsuperscript{nd}, 2017 observations of the NGC~1275 flare. The best fit model of the joint likelihood without \acp{alp} (with an \ac{alp} of ${\axm = 1.61\times 10^{-8} \eV}$ and ${\axg = 3.16 \gunit}$) is shown as a solid (dashed) black line for one realization of the \ac{icmf}.%
    }
    \label{fig:ngc1275_alp_sed}
\end{figure}

Since \pkg{gammapy} forward folds the model spectrum with the detector response to yield the predicted counts from the source, bin-by-bin correlations are self-consistently accounted for. As a result, it is possible to bin the \acs{vts} data below the level of the instrument's energy resolution\textemdash small enough to resolve \alpphot spectral  features\textemdash albeit, of course, with diminishing returns \cite{ajello_search_2016}. We have chosen 30 bins per decade ($\sim 15 - 25\%$ of the \acs{vts} energy resolution \cite{park_performance_2016}) for this analysis, as a compromise between resolution and computational overhead. The \ac{sed} for the flaring data is shown in \cref{fig:ngc1275_alp_sed}, where it is fit by the null hypothesis (no-\acs*{alp}) case modeled by \cref{eqn:nullhyp}. The overall best-fit model for the alternative hypothesis (\cref{eqn:alphyp}) is also shown.

\section{Statistical Methods}
\label{sec:statmethods}
The goal of this section is to describe the process towards yielding statistics which can:
\begin{itemize}
    \item Describe the likelihood of the existence of \acp{alp}
    \item Exclude regions of \ac{alp} parameter space
\end{itemize}
In order to yield both of these results, it will be necessary to scan $(\axm,~\axg)$-space in the regime where we expect \acs{vts} to provide the greatest sensitivity. Following from \cref{eqn:Ecrit}, a reasonable choice is to cover ${\axm \in [3,700] \neV}$ and ${\axg \in [0.1,10] \gunit}$. This range will allow $E_{\rm crit}^{\rm low}$ to fall in the energy range where \acs{vts} is most sensitive \cite{park_performance_2016}. Note that this parameter space is consistent and complementary with previous studies done with \Fermi-\acs{lat} on this source \cite{ajello_search_2016,cheng_revisiting_2021}. We choose to scan a $14\times9$ logarithmically spaced grid in $(\axm,~\axg)$-space to perform this study, with the goal of balancing computational overhead with resolution in the results.

\subsection{ALP Likelihood}\label{sec:ts}
The first question we can address is whether our results indicate evidence for the existence of \acp{alp}. We will seek to compare the likelihood of the most likely \ac{alp} case, i.e. the set of \ac{alp} parameters (\axmhat, \axghat) which best represent the observed data, with the maximum likelihood of the no-\acs*{alp} model (the null hypothesis). 

\begin{sloppypar}
This information is encapsulated in the test statistic:
\begin{equation}\label{eqn:ts}
    \TS = -2\ln\left(\frac%
        {\Lik_0 (\mu (\bm{{\Bar{\theta}}} ), {\Bar{b}} | \bm{D} )}
        % {\Lik_0(\bm{{\Bar{\theta}}})}%
        {\Lik_{\rm ALP}(\mu (\axmhat, \axghat, \bm{{\Hat{B}}_{95}}, \bm{{\Hat{\Hat{\theta}}}} ), \Hat{\Hat{b}} | \bm{D})}%
        %{\Lik_{\mathrm{ALP}}(\Hat{m}_a,\Hat{g}_{a\gamma},\bm{{\Hat{B}}}_{95},\bm{{\Hat{\Hat{\theta}}}})}%
    \right),
\end{equation}
a special case\footnote{This is \cref{eqn:lambda} for the case in which $(\axm, \axg) = \mathbf{0}$.} of the profile log-likelihood ratio test \cite{rolke_limits_2005}, which specifically measures the data's incompatibility with the null hypothesis as compared to the test hypothesis \cite{abdalla_sensitivity_2021}. ${\Lik_0 (\mu (\bm{{\Bar{\theta}}} ), {\Bar{b}} | \bm{D} )}$ is the maximum likelihood for the no-\acs*{alp} model, with profiled expected background ${\Bar{b}}$ and parameters of best-fit $\bm{\Bar{\theta}}$, and ${\Lik_{\rm ALP}(\mu (\axmhat, \axghat, \bm{{\Hat{B}}_{95}}, \bm{{\Hat{\Hat{\theta}}}} ), \Hat{\Hat{b}} | \bm{D})}$ is the maximum likelihood over the entire \ac{alp} grid (indicated by the additional hats). If we find the likelihood of the \ac{alp} case to be higher than that of the no-\acs*{alp} case, then this would be a potential indicator of the existence of the \ac{alp} case.
\end{sloppypar}

However, it is not a trivial exercise to quantify the significance of this \TS. Due to the nonlinearity in the relationship between the spectral oscillations and \ac{alp} parameters, as well as the degeneracy between $\axg$ and magnetic field strength in \cref{eqn:oscprob,eqn:deltaosc}, one can not simply convert \TS value into a significance following the usual Wilks's Theorem \cite{wilks_large-sample_1938}. As such, the null distribution is \textit{a priori} unknown, and must be derived from \ac{mc} simulations.
\info[inline]{``For example, one prerequisite stipulates that two distinct points within the parameter space should yield two unique predictions. Unfortunately, this condition does not hold up when considering values of the couplings \axg close to zero (e. g., there is no ALP effect). In such cases, any variation in the mass ma will inevitably lead to identical predictions, thus violating this essential criterion. Therefore assuming a $\chi^2$-distribution with two degrees of freedom for the statistic T S(\axg, \axm) would lead to a wrong coverage'' from \textcite{batkovic_searches_2024}.}

We perform \ac{mc} simulations of the data assuming the no-\acs*{alp} case, and examine the \TS distribution to establish the \acp{cl} at which one could reject the no-\acs*{alp} hypothesis. Simulations were generated starting from the best-fit null hypothesis model to the data. To do so, we convolve the null hypothesis model best fit to the observed data with the \acp{irf} to make predictions for the number of counts. Then Poisson random numbers are generated, based on the expected number of events, to yield simulated counts. In a similar vein to the discussion in \cref{sec:dataana}, the lack of observed background counts in the highest-energy bins will bias the high-energy bins of simulations produced using the observed data. To account for this, we model the background counts above $E_{\rm thresh, bkg} = 1 {\rm~TeV}$ as a power law with index 3, chosen to closely match the background data at lower energies with higher statistics. The power law is normalized such that, integrated above $E_{\rm thresh, bkg}$, it yields the total number of measured background counts above $E_{\rm thresh, bkg}$ \cite{jouvin_statistical_2016}\footnote{Note that this procedure is only applied to the data which is used to produce the simulations, not the simulations themselves.}.

\begin{figure}
    \centering
    \includegraphics[width=\columnwidth]{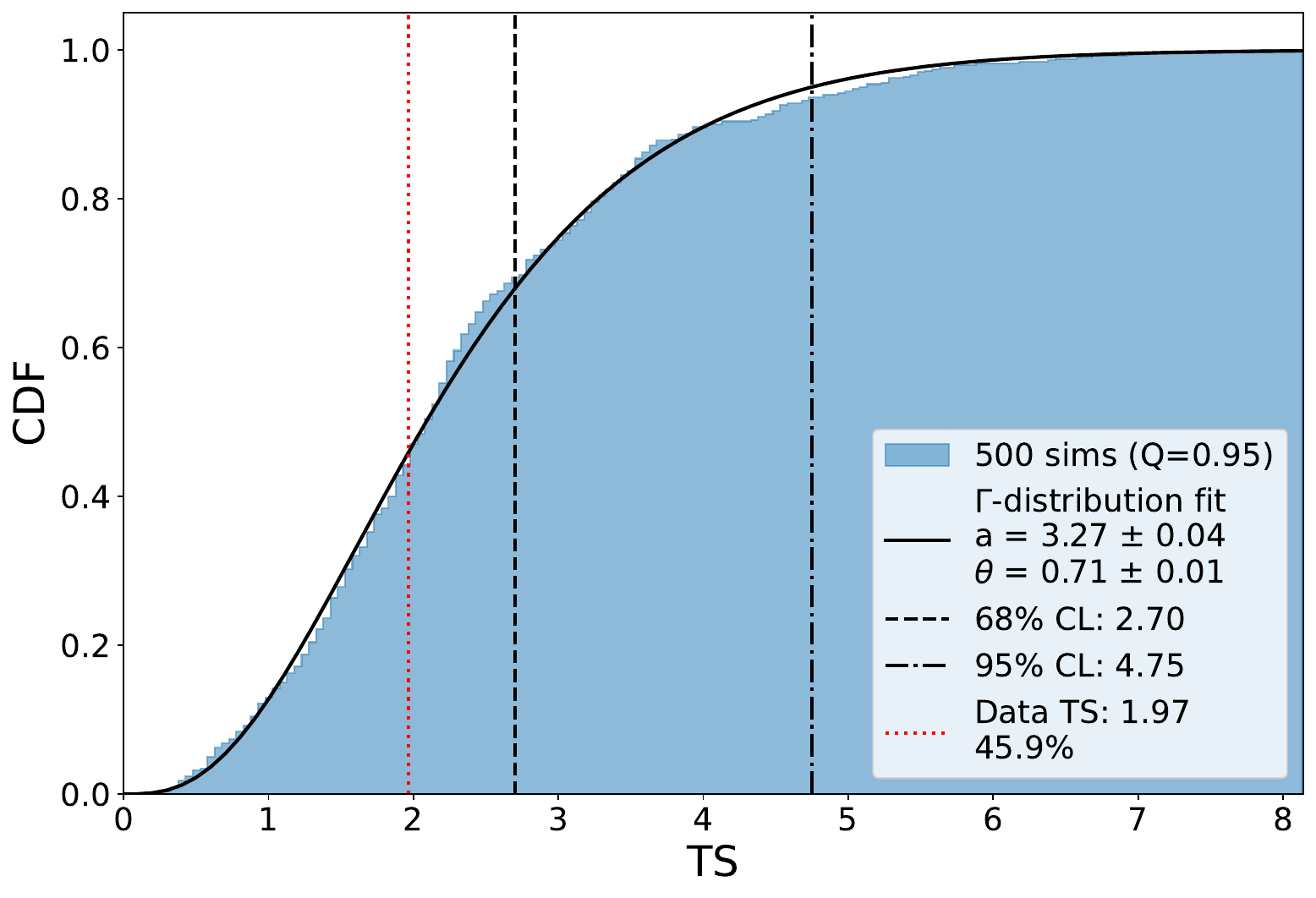}
    \caption[\Acl{cdf} for the \TS values calculated from 500 simulations produced assuming the no-\acs*{alp} case.]{\Acf{cdf} for the \TS values calculated from 500 simulations produced assuming the no-\acs*{alp} case. The \acs{cdf} is fitted with a modified $\Gamma$ distribution. Black vertical lines show the 68\% and 95\% \TS thresholds, and the red dotted line shows the \TS value of the data, with its associated \ac{cl} shown in the legend.}
    \label{fig:TS_dist}
\end{figure}

A total of 500 simulations are generated, each of which is re-analyzed following the exact procedure described before, yielding a \TS distribution from the fits to the simulations, shown in \cref{fig:TS_dist}. Following from \textcite{abdalla_sensitivity_2021}, we fit the \TS distribution with a modified $\Gamma$ distribution, from which we can extract the threshold values at 95\% confidence. From the best fit modified $\Gamma$ distribution, a $\TS > 4.75$ is required to reject the no-\acs*{alp} hypothesis with 95\% confidence.

\subsection{Exclusion Regions of ALP Parameter Space}\label{sec:excl}
Regardless of whether any indication of \acp{alp} is present, it is possible to exclude regions of the $(\axm, \axg)$ parameter space which are inconsistent with the observations. This has historically been done with the generalized profile log-likelihood ratio test \cite{rolke_limits_2005}, see, e.g., Ref.\@~\cite{ abdalla_sensitivity_2021}:
\begin{equation}\label{eqn:lambda}
    \lambda(\axm, \axg) = -2\ln\left(\frac%
        {\Lik_{\rm ALP}(\mu (\axm, \axg, \bm{{B_{95}}}, \bm{{\Hat{\theta}}} ), \Hat{b} | \bm{D})}
        % {\Lik_{\rm ALP}(\axm,\axg,\bm{B}_{95},\boldsymbol{\Hat{\theta}})}%
        {\Lik_{\rm ALP}(\mu (\axmhat, \axghat, \bm{{\Hat{B}}_{95}}, \bm{{\Hat{\Hat{\theta}}}} ), \Hat{\Hat{b}} | \bm{D})}
        % {\Lik_{\mathrm{ALP}}(\Hat{m}_a,\Hat{g}_{a\gamma},\bm{{\Hat{B}}}_{95},\bm{{\Hat{\Hat{\theta}}}})}%
    \right),
\end{equation}
which compares the best fit at each $(\axm, \axg)$ point in the grid with the overall best fit among the test points. Larger values of $\lambda$ will indicate that the best fit at the point under consideration is less representative of the data than the best fit overall.

A key component of the likelihood ratio test is that it is dependent on an unrestricted maximization (over the entire parameter space) \cite{casella_statistical_2002}. In the presence of any systematic uncertainties (for example, in energy resolution, \aclp{irf}, modeling) that are unaccounted for in the \ac{mle}, and especially in the generation of simulations, this unrestricted maximization may encounter (marginal) false positives, also known as Type I errors. Since the exclusion regions will be tied, by the denominator of \cref{eqn:lambda}, to the strongest signal found in the parameter space, such anomalous results would consequently result in artificially inflated exclusion \acp{cl}, which may extend exclusion regions to large parts of the unrestricted parameter space, including in regimes where the search ostensibly has no sensitivity \cite{read_modified_2000}. In the case of a true (marginal) positive signal, this is the desired behavior; however, as the truth case is unknown, the possibility for this inflationary effect motivates a more conservative approach.
%\footnote{A false positive with a one-tailed Gaussian significance $\gtrsim 1.64\sigma$ ($\gtrsim 1.96\sigma$, as usual, for the two-tailed Gaussian) is enough to exclude the null hypothesis at or above 95\% confidence.}.

The so-called ${\rm CL_s}$ method is a well-established tool for searches in high-energy experiments, which normalizes the test statistic used for exclusion with respect to the null hypothesis, rather than the supremum of the space as in \cref{eqn:lambda}, and has an additional renormalization which reduces the risk of constraining regimes where a search has no sensitivity \cite{read_modified_2000, read_presentation_2002, workman_review_2022}. In light of these advantages, we opt to utilize the ${\rm CL_s}$ method instead. Recent articles pursuing this type of study have also adopted the ${\rm CL_s}$ method for the same reasons \cite{gao_constraints_2024, gao_constraints_2024-1}.

To do so, we can generalize the test statistic in \cref{eqn:ts} to:
\begin{equation}\label{eqn:qstat}
    Q(\axm, \axg) = -2\ln\left(\frac%
        {\Lik_0 (\mu (\bm{{\Bar{\theta}}} ), {\Bar{b}} | \bm{D} )}
        % {\Lik_0(\bm{{\Bar{\theta}}})}%
        {\Lik_{\rm ALP}(\mu (\axm, \axg, \bm{{B}_{95}}, \bm{{\Hat{\theta}}} ), \Hat{b} | \bm{D})}%
        %{\Lik_{\mathrm{ALP}}(\Hat{m}_a,\Hat{g}_{a\gamma},\bm{{\Hat{B}}}_{95},\bm{{\Hat{\Hat{\theta}}}})}%
    \right),
\end{equation}
where again the numerator is the maximum likelihood under the null hypothesis, with profiled expected background ${\Bar{b}}$ and parameters of best-fit $\bm{\Bar{\theta}}$. The denominator now, however, is no longer profiled over the unrestricted parameter space, and is rather the maximum likelihood under the alternative hypothesis for a given fixed ${(\axm, \axg)}$. We can thus systematically scan the parameter space, finding a $Q_{\rm obs}(\axm, \axg)$ for the observed data at each test point, which can then be used to assess the strength of exclusion at that point. 

For some example point $(\axm, \axg)$, we seek to understand $Q_{\rm obs}(\axm, \axg)$ in the context of the underlying distribution of $Q(\axm, \axg)$ in the cases where the null hypothesis is true and where the alternative hypothesis is true, respectively. From \cref{sec:ts}, 500 simulations representing the null hypothesis truth case have already been produced. Each of these simulations was analyzed as described in \cref{sec:dataana}, the results from which were passed through \cref{eqn:qstat}, yielding a distribution $\{Q(\axm, \axg)\}_{0}$ for each (\axm, \axg) pair tested that can again be fitted with a modified $\Gamma$ distribution. Additionally, we now also simulate the injected \ac{alp} signal at each point in the parameter space to represent the alternative hypothesis truth case, with 500 simulations produced at each point. Following the same analysis, we find a distribution $\{Q(\axm, \axg)\}_{\rm ALP}$.
Two example distributions of $\{Q(\axm, \axg)\}_{0}$ and $\{Q(\axm, \axg)\}_{\rm ALP}$ along with the corresponding $Q_{\rm obs}(\axm, \axg)$ are shown in Appendix~\ref{sec:appendixA}. 

The ${\rm CL_s}$ value is colloquially defined as ${{\rm CL_s} = p_{\mu}/(1-p_{b})}$ \cite{workman_review_2022}, where $\mu$ and $b$ refer to the signal and background-only models respectively. Applied in this context, with the alternative hypothesis $H_{\rm ALP}$ and null hypothesis $H_{0}$, we define the ${\rm CL_s}$ value as:
\begin{linenomath*}
\begin{align}\label{eqn:cls}
\begin{split}
    {\rm CL_s}(\axm, \axg) &= \frac{p_{\rm ALP}}{1-p_{0}} \\
    &= \frac{P({\rm reject~}H_{\rm ALP} | H_{\rm ALP} = {\rm TRUE})}{P( {\rm accept~}H_{\rm 0} | H_{\rm 0} = {\rm TRUE})}.
    % &= \frac{P(Q_{\rm ALP} \leq Q_{\rm obs}(\axm, \axg))}{P(Q_{\rm 0} \leq Q_{\rm obs}(\axm, \axg))}.
\end{split}
\end{align}
\end{linenomath*}
Thus, the ${\rm CL_s}$ value can be interpreted as a p-value that has been normalized to the acceptance probability for $H_0$. As a result, this means that the less separation that exists between the distributions $\{Q(\axm, \axg)\}_{\rm ALP}$ and $\{Q(\axm, \axg)\}_{\rm 0}$ (i.e. the less sensitive we are in our search), the more severely penalized our ${\rm CL_s}$ value will be. This is seen in that $p_{\rm ALP}$ is the strict lower bound of the ${\rm CL_s}$ value. When the ${\rm CL_s}$ value is smaller than some number $\alpha$, say for example $\alpha=0.05$, the test point is considered incompatible with the observed data at the $1 - \alpha$ \ac{cl}; in our example, this would correspond to an exclusion at 95\% confidence.
Both the numerator and denominator of \cref{eqn:cls} are computed simply as the \acp{cdf} of the $\{Q(\axm, \axg)\}_{\rm ALP}$ and $\{Q(\axm, \axg)\}_{\rm 0}$ distributions respectively, evaluated at $Q_{\rm obs}(\axm, \axg)$. Denoting the \ac{cdf} from the simulation with an injected \ac{alp} signal at $(\axm, \axg)$ with $\mathcal{F}_{\axm, \axg}(Q(\axm, \axg))$ and without an ALP signal with $\mathcal{F}_0(Q(\axm, \axg))$, the ${\rm CL_s}$ therefore becomes
\begin{equation}
{\rm CL_s}(\axm, \axg) = \frac{\mathcal{F}_{\axm, \axg}(Q_\mathrm{obs}(m_a, g_{a\gamma}))}{\mathcal{F}_0(Q_\mathrm{obs}(\axm, \axg))}.    
\end{equation}

\section{Results and Discussion}
\label{sec:alp-results}
We find in our results no evidence for the existence of \acp{alp}. Subsequently, we exclude portions of the \ac{alp} mass and coupling parameter space at up to the 80\% \ac{cl}.

The test statistic for the observed data was measured as $\TS=1.97$, found at the test point ${(\axm = 1.61\times 10^{-8} \eV}$, ${\axg = 3.16 \gunit)}$. This result is shown as the red dotted line which has been included in \cref{fig:TS_dist}, which indicates a data \TS which lies just below the 50\% quantile of the modified $\Gamma$-distribution fit to the distribution. 
% https://www.ncl.ac.uk/webtemplate/ask-assets/external/maths-resources/statistics/hypothesis-testing/one-tailed-and-two-tailed-tests.html#Worked%20Example%203
This \ac{cl} can be converted to the one-tailed Gaussian equivalent standard deviation $\sigma$ through the inverse of the complementary error function \cite{workman_review_2022}: ${\sigma = -\sqrt{2}\erfc^{-1}(2 \cdot {\rm CL})}$, corresponding to a rejection significance of the null hypothesis of $-0.1 \sigma$\footnote{The equivalent significance of the two-tailed Gaussian, which has been presented in other related works, is given by ${\sigma = \sqrt{2} \erf^{-1}({\rm CL})}$, where $\erf^{-1}$ is the inverse of the error function. It yields a two-tailed equivalent Gaussian significance of $0.61 \sigma$, which remains consistent with the conclusions drawn from the one-tailed Gaussian significance.}. This indicates no evidence for a rejection of the null hypothesis in favor of the alternative hypothesis.

\begin{figure*}
    \centering
    \includegraphics[width=\textwidth]{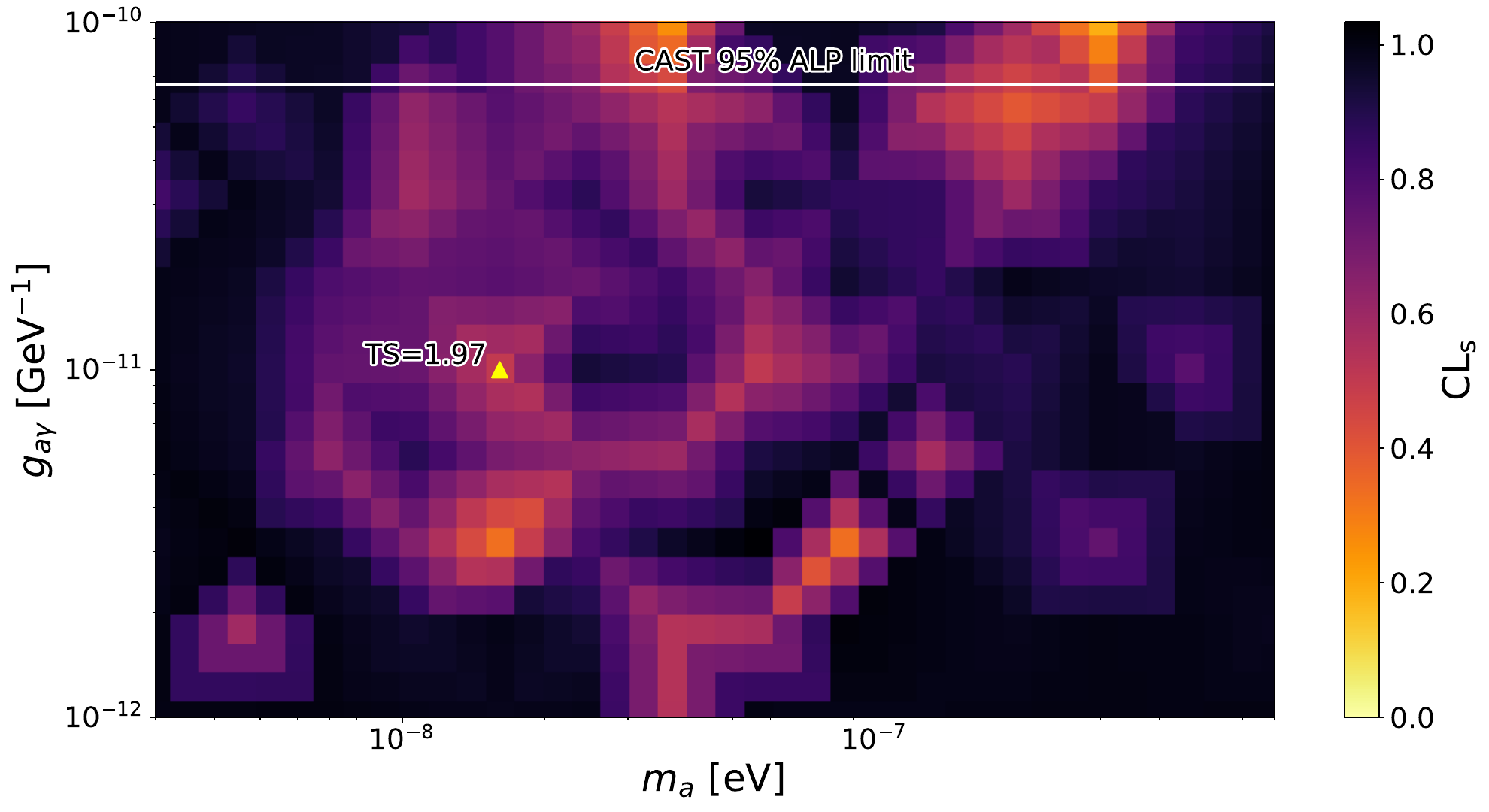}
    \caption[Map of \acs*{alp} parameter ${\rm CL_s}$ values.]{Map of \acs*{alp} parameter ${\rm CL_s}$ values. Points are excluded with confidence $1-{\rm CL_s}$. The point of highest \acs{vts} \TS is indicated by a yellow triangle and annotated with its value. As a reference, 95\% exclusions from \acs*{cast} are above the white line. For visualization purposes, the sparse $14 \times 9$ grid has been upsampled by a factor of three and smoothed with linear interpolation.}
    \label{fig:cls_space}
\end{figure*}

From the ${\rm CL_s}$ values, we can still examine the ability of our search to constrain the region of \ac{alp} parameter space tested. This is done as described in \cref{sec:excl}, and is shown in \cref{fig:cls_space}. \rev{We note that \cref{fig:cls_space} has been upsampled by a factor of three and smoothed with linear interpolation for visualization purposes}. 
As can be seen, the region of the tested space excluded with the highest confidence is where ${\axg \gtrsim 3 \gunit}$ in the range ${2 \times 10^{-7} {\rm~eV} \lesssim \axm \lesssim 4 \times 10^{-7} {\rm~eV}}$, peaking at $\sim 80\%$ confidence exclusion. As our constraints do not anywhere reach the 95\% \ac{cl}, our results with the ${\rm CL_s}$ method are not considered constraining. Nonetheless, our results show no inconsistencies with exclusion regions found in previous studies \cite[e.g.,][]{abe_constraints_2024, davies_constraints_2023, li_limits_2021}.
%Our results, presented alongside the current state of the field, are shown in \cref{fig:allexcls}.

As a cross-check, we also conducted the analysis using the profile likelihood method described at the start of \cref{sec:excl}. This method yielded an exclusion map characterized by the same overall trend, with constraints not exceeding the 95\% \ac{cl} anywhere in the parameter space. However, all of the strongest constraints were found to be tighter in the profile likelihood case, emphasizing the less conservative nature of this approach (see Appendix~\ref{sec:appendixB}). Marked differences between the two methods do appear near the most likely \ac{alp} test point in the parameter space (corresponding to $\TS=1.97$), highlighting one of the key differences between the two approaches\textemdash the regime nearest to the most likely \ac{alp} point will always be minimally constrained in the profile likelihood approach, whereas the constraints set by the ${\rm CL_s}$ method will be insensitive to this feature.

Our results are considerably less constraining than those shown in the analysis of the same source, during the same flare, with \ac{magic} in \textcite{abe_constraints_2024}. We note as a potential interpretation for this discrepancy the differences in datasets, instruments, and methodology. \Ac{magic} observed the target's flaring state a day before \ac{vts}, recording a flux $\sim 3 \times$~higher than that measured by \ac{vts} the following day, yielding a more robust \ac{s/n}. Additionally, post-flare and quiescent state data were included in that work. We also note the presence of two points in the space examined by the \ac{magic} analysis, ${(\axm = 1.00 \times 10^{-7} {\rm~eV},}$ ${\axg= 2.71 \times 10^{-10} {\rm~GeV^{-1}})}$ and ${(\axm = 2.15 \times 10^{-8} {\rm~eV},}$ ${\axg= 3.81 \times 10^{-12} {\rm~GeV^{-1}})}$, which reject the null hypothesis at $\gtrsim 95\%$ confidence in favor of the alternative hypothesis. Both points lie in regimes that have been previously excluded at or above 95\% confidence by other searches \cite{cast_collaboration_new_2017, noordhuis_novel_2023}. As the profile likelihood was utilized in this case, it is likely that the presence of these marginally significant points contributed to the depth and extent of their exclusion regions, as discussed in \cref{sec:excl}.\unsure{This is a potential knock to the legitimacy of the MAGIC results, should be careful how we frame this. We may want to share this excerpt with the MAGIC collaboration prior to submission to initiate a conversation.}

In considering future applications for this analysis, the advent of the \acf{ctao} offers improved prospects for constraining the \ac{alp} parameter space. Notably, \ac{ctao}'s exceptional energy resolution means an increased sensitivity to the predicted small-scale spectral irregularities induced by \ac{alp} oscillations in the presence of external magnetic fields. Further, the two-array design of \ac{ctao} will continue to provide the nearly full-sky \ac{vhe} coverage offered by existing \ac{iactel} arrays, meaning uninterrupted access to a plethora of potential sources residing in galaxy clusters that could be probed. A preliminary look at the \ac{ctao} capabilities was investigated on the same source, NGC~1275, in \textcite{abdalla_sensitivity_2021}, which underscores the importance of using flaring data, as well as the value of improved energy resolution in next-generation instruments. The study also highlights the impact of systematic uncertainties in model assumptions and detector response. For the model assumptions, of particular focus were magnetic field strength, the index of magnetic field turbulence, the minimum turbulence scale, and the power law index $\eta$ relating electron density with magnetic field strength. It shows that limits would be expected to weaken as any one of these parameters decreases (with the exception of $\eta$, which would yield weaker limits as it increases). When systematic uncertainties in the instrument responses were included, limits were also found to degrade. These findings apply to our study as well, and are therefore an important caveat to consider when appraising our results.

\section{Conclusion}
\label{sec:conclusion}
Although they may not solve the strong \ac{cp} problem like the axion, \acp{alp} are still a compelling target for experimental searches for a number of reasons. Principal among these is that \acp{alp} are theoretically well-motivated, emerging naturally from a number of string theories and other extensions to the \ac{sm}. The large parameter space implied by the relaxation of the $\axm/\axg$ relationship in going from axions to \acp{alp} allows for a diverse set of experimental approaches that may be employed to provide \ac{alp} constraints. In particular, X- and \gray{} observations of astrophysical objects, especially those embedded in strong magnetic fields over large spatial scales, have proven to be a useful tool in constraining the region of \ac{alp} parameter space for masses ${\axm \lesssim 100~{\rm neV}}$ \citep[e.g.,][]{batkovic_axion-like_2021,abramowski_constraints_2013, ajello_search_2016, reynolds_astrophysical_2020, Buehler2020a, matthews_how_2022, sisk-reynes_new_2021, davies_constraints_2023, abe_constraints_2024}. \acs{vts}, with its \ac{vhe} observations of dramatic \acp{agn} flares, is well positioned to contribute in this regime in accordance with \cref{eqn:Ecrit}.
%to high enough masses where \acp{alp} could comprise all of the \ac{dm} in the Universe (noting that the precise cut-off for that qualification is framework-dependent) \cite{paola_arias_wispy_2012, abdalla_sensitivity_2021}.

Motivated by these prospects, we have used the \acs{vts} spectrum of the \acl*{rg} NGC~1275 during a major flare to search for the imprint of \acp{alp}. In particular, we exploit the turbulent magnetic field of the Perseus cluster, in which NGC~1275 is embedded, as the main mixing region for the \alpphot beam. The large spatial scale of the cluster, along with its considerable ${\mathcal{O}(10~{\rm \mu G})}$ magnetic field strength, makes it a particularly promising target for this technique. 
%In considering some of the pitfalls associated with the commonly used profile likelihood method, we adopted a more conservative approach to determining the confidence of exclusion regions, the ${\rm CL_s}$ method, the first time it has been used to set constraints using the Perseus cluster. 
In considering some of the pitfalls associated with the commonly used profile likelihood method, we adopted the ${\rm CL_s}$ method. This is a more conservative approach to determining the confidence of exclusion regions, and our work is the first time it has been used to set constraints with the Perseus cluster.
With this approach, our analysis yielded conservative constraints, peaking at an 80\% confidence exclusion at ${(\axm = 3.03 \times 10^{-7} {\rm~eV},}$ ${\axg= 1.00 \times 10^{-10} {\rm~GeV^{-1}})}$, and did not contradict earlier measurements. In future studies of this type, we advocate for the continued adoption of the ${\rm CL_s}$ method considering the limitations of the profile likelihood approach.

% Our analysis yielded a small area excluded at 68\% confidence in the $(\axm, \axg)$ parameter space, 
% excluding a region in agreement with and complementary to the constraints set by other searches using similar techniques (see \cref{fig:allexcls})
% roughly excluding masses ${50 \lesssim \axm \lesssim 300~{\rm neV}}$ for ${\axg \gtrsim 0.3 \times 10^{-11} \gunit}$ (details shown in \cref{fig:lambda_contours} and \cref{fig:allexcls}).
% placing constraints on \axg over the mass range ${200 \lesssim \axm \lesssim 350 {\rm~neV}}$, excluding ${\axg \gtrsim 3 \gunit}$ (details shown in \cref{fig:lambda_contours} and \cref{fig:allexcls}).

The next-generation \ac{ctao}, with its improved energy resolution and extensive energy sensitivity, is expected to further the capabilities of \acp{iactel} to contribute meaningful constraints in these types of analyses, bringing this technique into a new era of \gray astronomy and astrophysics.

\begin{acknowledgments}
This research is supported by grants from the U.S. Department of Energy Office of Science, the U.S. National Science Foundation and the Smithsonian Institution, by NSERC in Canada, and by the Helmholtz Association in Germany. This research used resources provided by the Open Science Grid, which is supported by the National Science Foundation and the U.S. Department of Energy's Office of Science, and resources of the National Energy Research Scientific Computing Center (NERSC), a U.S. Department of Energy Office of Science User Facility operated under Contract No. DE-AC02-05CH11231. We acknowledge the excellent work of the technical support staff at the Fred Lawrence Whipple Observatory and at the collaborating institutions in the construction and operation of the instrument.

C. A. was supported by NSF grant PHY-2110497. Q.~F. acknowledges the support from NSF grant PHY-2411860. 
M.~M. acknowledges the support from the Deutsche Forschungsgemeinschaft (DFG, German Research Foundation) under Germany's Excellence Strategy – EXC 2121 ``Quantum Universe'' – 390833306 and from the European Research Council (ERC) under the European Union’s Horizon 2020 research and innovation program Grant agreement No. 948689 (AxionDM).
\end{acknowledgments}

\appendix

\section{Test statistic distributions from the ${\rm CL_s}$ method}
\label{sec:appendixA}

\begin{figure}
    \centering
    \includegraphics[width=\columnwidth]{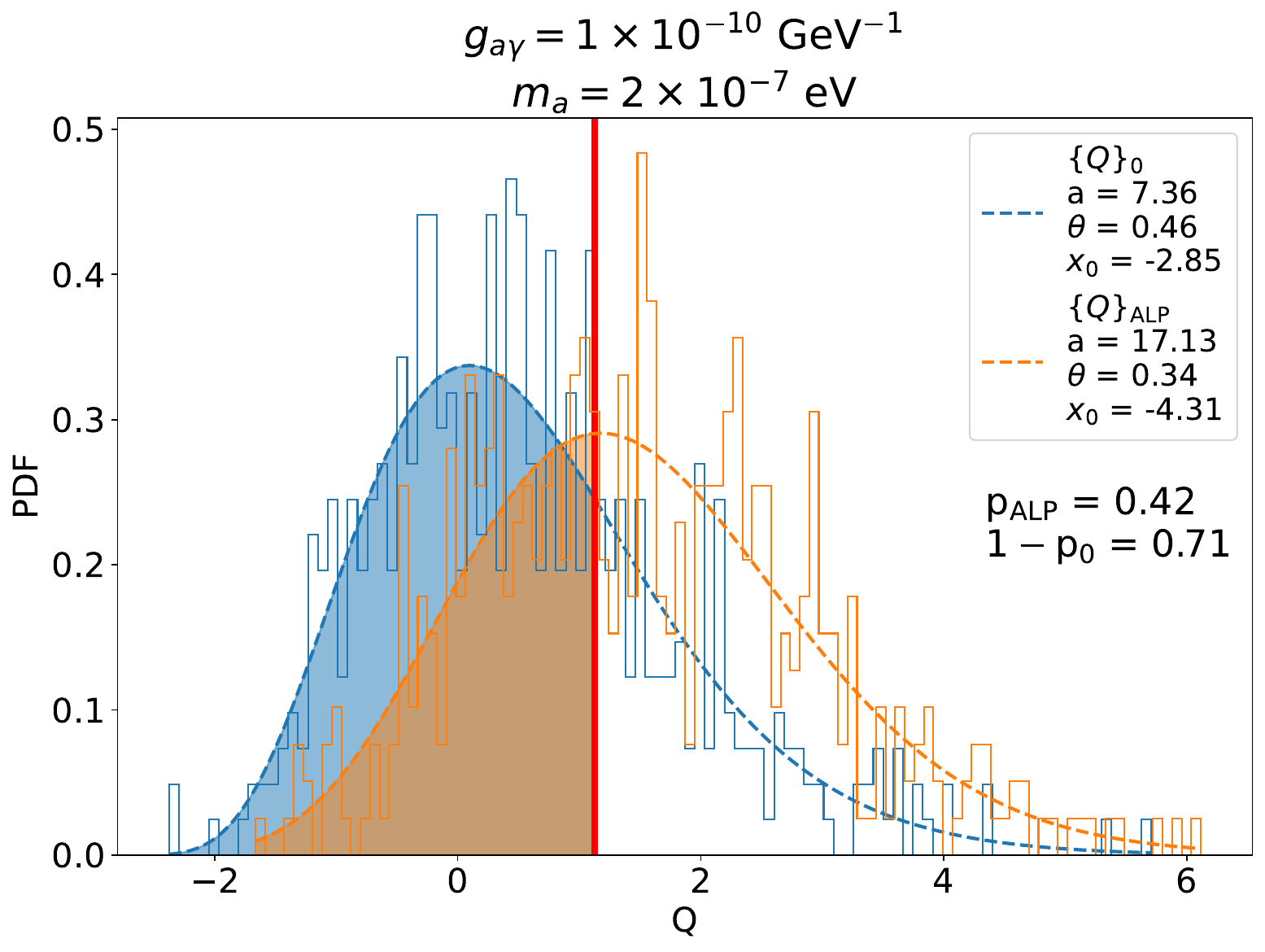}
    \includegraphics[width=\columnwidth]{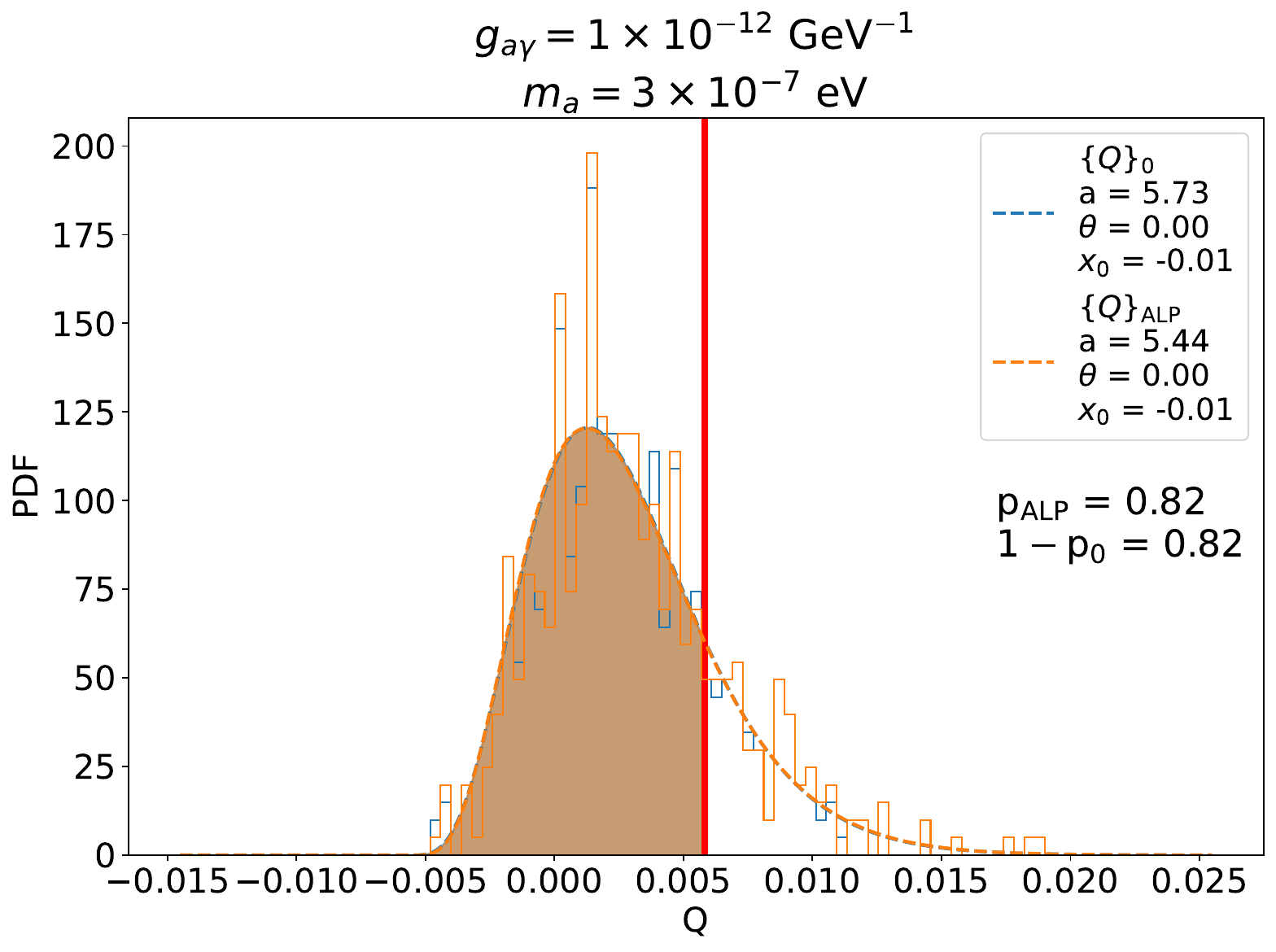}
    \caption[Qdist]{The distribution of the test statistic ($Q$) values calculated using Equation~\ref{eqn:qstat} at two test points ${(\axm = 2\times 10^{-7} \eV}$, ${\axg = 1 \ensuremath{{\rm~\times~10^{-10}~GeV^{-1}}}\xspace)}$ and ${(\axm = 3\times 10^{-7} \eV}$, ${\axg = 1 \ensuremath{{\rm~\times~10^{-12}~GeV^{-1}}}\xspace)}$, respectively. The blue distributions are for the null hypothesis, and the orange distributions are for the alternative ALP hypothesis. The red vertical line is the test statistic $Q$ calculated from the data. $a$, $\theta$, and $x_0$ are the shape, scale, and location parameters resulting from a fit to a gamma function.}
    \label{fig:qstat_dist}
\end{figure}

To illustrate the performance of the ${\rm CL_s}$ method, Fig.~\ref{fig:qstat_dist} shows example distributions of the test statistic ($Q$) under both the null and ALP hypotheses at two test points ${(\axm = 2\times 10^{-7} \eV}$, ${\axg = 1 \gunit)}$ and ${(\axm = 3\times 10^{-7} \eV}$, ${\axg = 1 \ensuremath{{\rm~\times~10^{-12}~GeV^{-1}}}\xspace)}$. At each test point, including the test points not shown in Fig.~\ref{fig:qstat_dist}, a test statistic distribution for the null hypothesis ($\{Q\}_0$) is calculated with the 500 null simulations as described in Sec.~\ref{sec:ts} (shown in blue in Fig.~\ref{fig:qstat_dist}). Similarly, a test statistic distribution for the alternative hypothesis ($\{Q\}_\mathrm{ALP}$) is calculated using the 500 simulations with the injected ALP signal at each test point as described in Sec.~\ref{sec:excl} (shown in orange in Fig.~\ref{fig:qstat_dist}). 

When an instrument lacks sensitivity to an ALP signature in a certain region of the mass and coupling strength parameter space (e.g., Fig.~\ref{fig:qstat_dist} bottom panel; ${(\axm = 3\times 10^{-7} \eV}$, ${\axg = 1 \ensuremath{{\rm~\times~10^{-12}~GeV^{-1}}}\xspace)}$), the null and alternative distributions are similar, leading to a ${\rm CL_s}$ close to 1 (see Eq.~\ref{eqn:cls}). A ${\rm CL_s}$ value of nearly 1 indicates low confidence in excluding the ALP hypothesis with the given parameters. Hence, the ${\rm CL_s}$ method does not exclude parameter space where the instrument has low sensitivity to the signal being searched for. 

At the other test point (Fig.~\ref{fig:qstat_dist} top panel; ${(\axm = 2\times 10^{-7} \eV}$, ${\axg = 1 \gunit)}$), the null and alternative distributions are clearly different, translating into a relatively low ${\rm CL_s}$ value (${p_{\rm ALP}}/{(1-p_{0})}\approx 0.6$) and some confidence to exclude the specific mass and coupling strength values. 

\section{Exclusion regions using the profile likelihood method}
\label{sec:appendixB}

Although the ${\rm CL_s}$ method is chosen as the analysis method of this work, we include in this appendix the exclusion regions obtained using the profile likelihood method described in Sec.~\ref{sec:excl}. The confidence level of excluding a given combination of the ALP mass and coupling strength is derived from the profile likelihood $\lambda$, as defined in Eq.~\ref{eqn:lambda}. 

\begin{figure}[hbt!]
    \centering
    \includegraphics[width=\columnwidth]{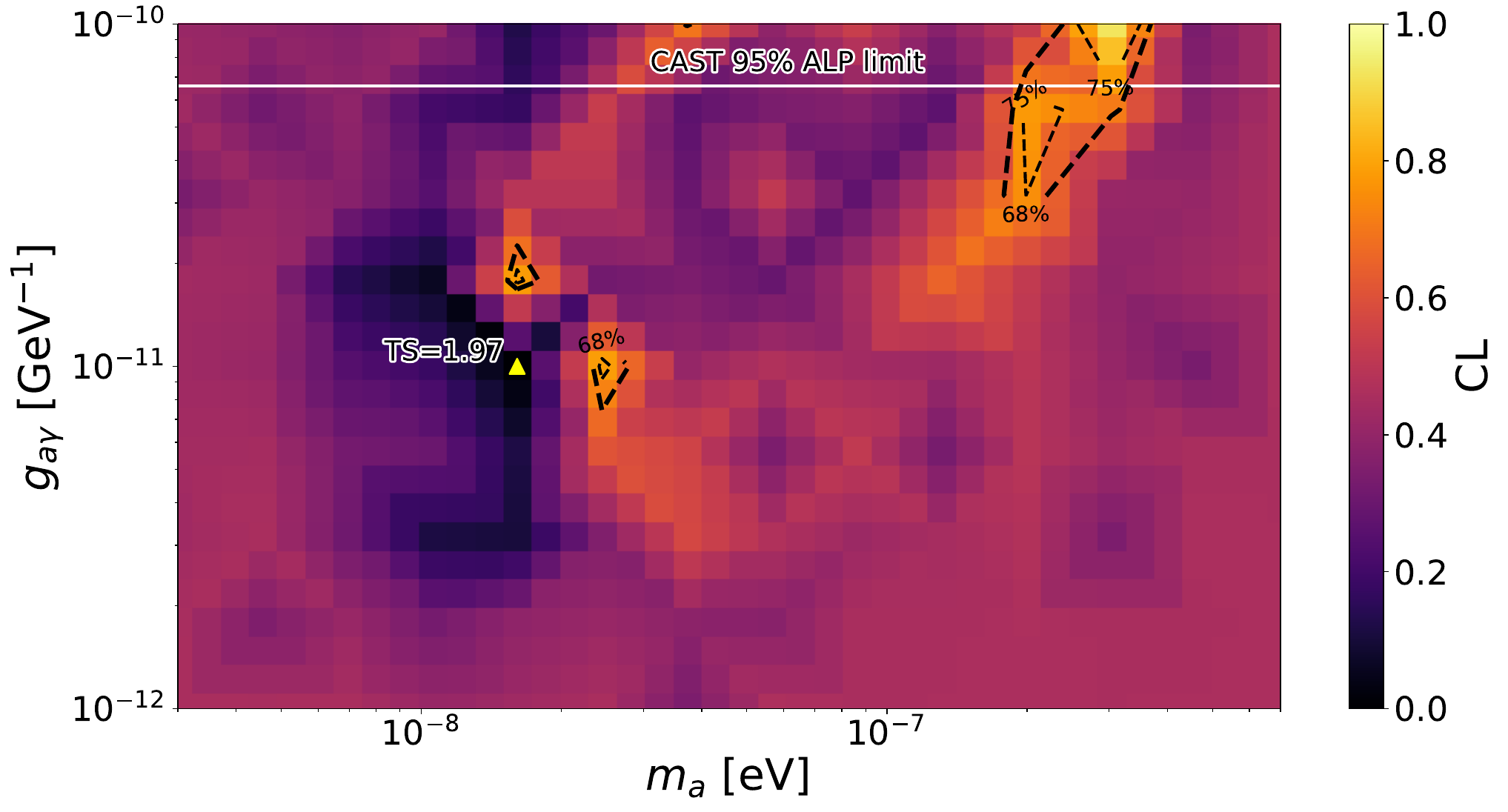}
    \caption[Lambda map]{The map of exclusion confidence levels derived from the $\lambda$ values calculated using Equation~\ref{eqn:lambda}, covering the same \acs*{alp} parameter space as Figure~\ref{fig:cls_space}. The dashed black contours indicate the regions excluded at 68\% and 75\%. The point of highest \acs{vts} \TS is indicated by a yellow triangle and annotated with its value.}
    \label{fig:profile_exclusion}
\end{figure}

The region excluded at 68\% and 75\% confidence levels using the profile likelihood method is consistent with similar previous studies. The exclusion region also matches the expectation for where VERITAS would be most sensitive to ALP/photon oscillation effects. However, the confidence of the exclusion is much higher compared to the \textbf{${\rm CL_s}$} method, since the exclusion confidence is derived relative to the overall best fit $(\axmhat, \axghat)$ among the test points.

\section{\rev{Map of \acs*{alp} parameter ${\rm CL_s}$ values without upsampling}}
\label{sec:appendixC}
\begin{figure}[hbt!]
    \centering
    \includegraphics[width=\columnwidth]{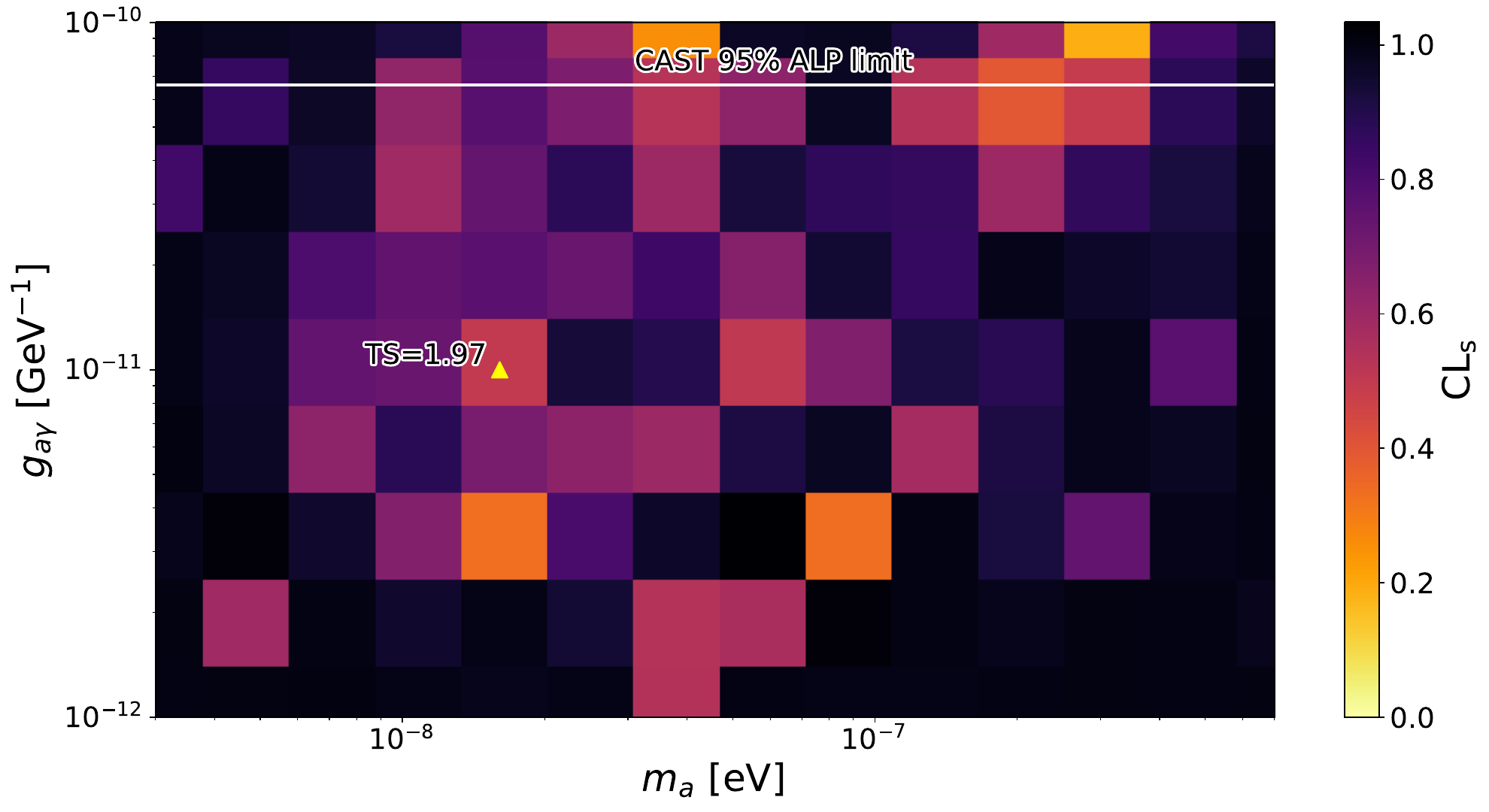}
    \caption[Map of \acs*{alp} parameter ${\rm CL_s}$ values without upsampling.]{\rev{Map of \acs*{alp} parameter ${\rm CL_s}$ values, similar to \cref{fig:cls_space}, but without any upsampling/smoothing. }}
    \label{fig:noupsamp_CLsMap}
\end{figure}

\rev{\cref{fig:cls_space} is shown with an upsampling by a factor of three with linear interpolation for visualization purposes. For completeness, we show the raw map of \acs*{alp} parameter ${\rm CL_s}$ values without any smoothing in \cref{fig:noupsamp_CLsMap}. }

\bibliographystyle{apsrev4-2}
\bibliography{references,extra}%qi_bib}% Produces the bibliography via BibTeX.

\end{document}

%% file: acro_local.tex
\DeclareAcronym{1d}{
short=1D,
long=one-dimensional,
tag=common,
}

\DeclareAcronym{2d}{
short=2D,
long=two-dimensional,
tag=common,
}

\DeclareAcronym{3d}{
short=3D,
long=three-dimensional,
tag=common,
}

\DeclareAcronym{4lac}{
short=4LAC,
long=fourth catalog of \acsp*{agn} detected by the \acs*{lat},
}

\DeclareAcronym{4fgl}{
short=4FGL,
long=fourth \Fermi-\acs*{lat} catalog
}

\DeclareAcronym{ac}{
short=AC,
long=alternating current,
tag=common,
}

\DeclareAcronym{adc}{
short=ADC,
long=analog-to-digital converter,
}

\DeclareAcronym{admx}{
short=ADMX,
long=Axion Dark Matter Experiment,
}

\DeclareAcronym{ajb}{
short=AJB,
long=Adler-Bell-Jackiw,
}

\DeclareAcronym{alp}{
short=ALP,
long=axion-like particle
}

\DeclareAcronym{alps}{
short=ALPS,
long=Any Light Particle Search,
}

\DeclareAcronym{agn}{
short=AGN,
short-plural=,
long=active galactic nucleus,
long-plural-form=active galactic nuclei,
list=active galactic nuclei
}

\DeclareAcronym{asas-sn}{
short=ASAS-SN,
long=All-Sky Automated Survey for Supernovae,
cite={kochanek_all-sky_2017},
}

\DeclareAcronym{asic}{
short=ASIC,
long=application-specific integrated circuit,
}

\DeclareAcronym{asl}{
short=a.s.l.,
long=above sea level,
}

\DeclareAcronym{astri}{
short=ASTRI,
long=Astrophysics with Mirrors via Italian Replication Technology,
cite={scuderi_astri_2019},
}

\DeclareAcronym{atel}{
short=ATel,
long=\textit{The Astronomer's Telegram},
}

\DeclareAcronym{atlas}{
short=ATLAS,
long=Asteroid Terrestrial-impact Last Alert System,
cite={tonry_atlas_2018}
}

\DeclareAcronym{bh}{
short=BH,
long=black hole,
}

\DeclareAcronym{bic}{
short=BIC,
long=Bayesian information criterion,
long-plural-form=Bayesian information criteria,
}

\DeclareAcronym{bllac}{
short=BL Lac,
long=BL Lacertae object,
}

\DeclareAcronym{blr}{
short=BLR,
long=broad-line region,
}

\DeclareAcronym{bns}{
short=BNS,
long=binary \acl{ns},
}

\DeclareAcronym{bz}{
short=BZ,
long=Blandford-Znajek,
}

\DeclareAcronym{cad}{
short=CAD,
long=computer-aided design,
}

\DeclareAcronym{casper}{
short=CASPEr,
long=Cosmic Axion Spin Precession Experiment,
}

\DeclareAcronym{cast}{
short=CAST,
long=\acs{cern} Axion Solar Telescope,
}

\DeclareAcronym{cat}{
short=CAT,
long=Cherenkov Array at Th\'{e}mis,
cite={barrau_cat_1998},
}

\DeclareAcronym{cbc}{
short=CBC,
long=compact binary coalescence,
}

\DeclareAcronym{ccd}{
short=CCD,
long=charge-coupled device,
}

\DeclareAcronym{cdf}{
short=CDF,
long=cumulative distribution function,
}

\DeclareAcronym{cdm}{
short=CDM,
long=cold \acl*{dm},
}

\DeclareAcronym{cern}{
short=CERN,
long=European Organization for Nuclear Research,
}

\DeclareAcronym{cfd}{
short=CFD,
long=constant-fraction discriminator,
}

\DeclareAcronym{chec}{
short=CHEC,
long=Compact High-Energy Camera,
cite={schoorlemmer_compact_2017},
}

\DeclareAcronym{cib}{
short=CIB,
long=cosmic \acl*{ir} background,
}

\DeclareAcronym{cl}{
short=CL,
long=confidence level,
first-style = long,
subsequent-style = long,
}

\DeclareAcronym{cmb}{
short=CMB,
long=cosmic microwave background,
}

\DeclareAcronym{cob}{
short=COB,
long=cosmic optical background,
}

\DeclareAcronym{cp}{
short=\ensuremath{CP},
sort=CP,
long=charge-parity
}

\DeclareAcronym{cr}{
short=CR,
long=cosmic ray,
}

\DeclareAcronym{ctao}{
short=CTAO,
long=Cherenkov Telescope Array Observatory,
}

\DeclareAcronym{dac}{
short=DAC,
long=digital-to-analog converter,
}

\DeclareAcronym{daq}{
short=DAQ,
alt=DACQ,
long=data acquisition,
}

\DeclareAcronym{dec}{
short=Dec.,
long=declination,
}

\DeclareAcronym{dc}{
short=DC,
sort=DC1,
long=Davies-Cotton,
}

\DeclareAcronym{dcurr}{
short=DC,
sort=DC2,
long=direct current,
tag=common,
}

\DeclareAcronym{dfsz}{
short=DFSZ,
long=Dine-Fischler-Srednicki-Zhitnisky,
}

\DeclareAcronym{dl}{
short=DL,
long=data~level,
}

\DeclareAcronym{dm}{
short=DM,
long=dark matter,
}

\DeclareAcronym{dr}{
short=DR,
long=dark run,
}

\DeclareAcronym{dsa}{
short=DSA,
long=diffusive shock acceleration,
cite={axford_acceleration_1977,krymskii_regular_1977,bell_accelerationI_1978,blandford_particle_1978},
}

\DeclareAcronym{ea}{
short=EA,
long=effective area,
}

\DeclareAcronym{eas}{
short=EAS,
long=extensive air shower
}

\DeclareAcronym{ebl}{
short=EBL,
long=extragalactic background light
}

\DeclareAcronym{ecpl}{
short=ECPL,
long=exponential cutoff power law,
}

\DeclareAcronym{ecsv}{
short=ECSV,
long=Enhanced Character-Separated Values,
}

\DeclareAcronym{ed}{
short=\pkg{evndisp},
sort=evndisp,
long=EventDisplay,
}

\DeclareAcronym{edm}{
short=eDM,
long=electric dipole moment,
}

\DeclareAcronym{eic}{
short=EIC,
long=external \acl*{ic}
}

\DeclareAcronym{em}{
short=EM,
long=electromagnetic
}

\DeclareAcronym{eos}{
short=EOS,
long=equation of state,
}

\DeclareAcronym{e/p}{
short=\ensuremath{e^{+}{\rm /}e^{-}},
long=electron/positron,
tag=convenience,
}

\DeclareAcronym{ew}{
short=EW,
long=electroweak
}

\DeclareAcronym{exosat}{
short=EXOSAT,
long=European X-ray Observatory Satellite,
}

\DeclareAcronym{fadc}{
short=FADC,
long=flash \acl*{adc},
}

\DeclareAcronym{fee}{
short=FEE,
long=front-end electronics,
}

\DeclareAcronym{fits}{
short=FITS,
long=Flexible Image Transport System,
}

\DeclareAcronym{flwo}{
short=FLWO,
long=Fred Lawrence Whipple Observatory,
}

\DeclareAcronym{fov}{
short=FoV,
long=field of view,
}

\DeclareAcronym{fps}{
short=FPS,
long=Filter Profile Service,
}

\DeclareAcronym{fri}{
short=FR-I,
long=Fanaroff-Riley Class I,
}

\DeclareAcronym{frii}{
short=FR-II,
long=Fanaroff-Riley Class II,
}

\DeclareAcronym{fsrq}{
short=FSRQ,
long=flat spectrum radio quasar,
}

\DeclareAcronym{gadf}{
short=GADF,
long=gamma-ray astronomy data format,
}

\DeclareAcronym{gbm}{
short=GBM,
long=Gamma-ray Burst Monitor,
cite={meegan_fermi_2009},
}

\DeclareAcronym{gct}{
short=GCT,
long=Gamma-ray Cherenkov Telescope,
cite={tibaldo_gamma-ray_2017},
}

\DeclareAcronym{grb}{
short=GRB,
long=gamma-ray burst,
}

\DeclareAcronym{gw}{
short=GW,
long=gravitational wave
}

\DeclareAcronym{gui}{
short=GUI,
long=graphical user interface,
}

\DeclareAcronym{gut}{
short=GUT,
short-plural=,
long=grand unification theory,
long-plural-form=grand unification theories,
}

\DeclareAcronym{hbl}{
short=HBL,
long=\ac{hsp} \acs*{bllac}
}

\DeclareAcronym{hdf}{
short=HDF,
long=Hierarchical Data Format,
}

\DeclareAcronym{he}{
short=HE,
long=high energy,
}

\DeclareAcronym{hegra}{
short=HEGRA,
long=High-Energy-Gamma-Ray Astronomy,
cite={daum_first_1997},
}

\DeclareAcronym{hess}{
short=H.E.S.S.,
long=High Energy Stereoscopic System,
cite={aharonian_observations_2006},
}

\DeclareAcronym{hew}{
short=HEW,
long=Heisenberg-Euler-Weisskopf,
}

\DeclareAcronym{hlv}{
short=HLV,
long={\acs*{ligo}-Hanford, \acs*{ligo}-Livingston and Virgo},
}

\DeclareAcronym{hsp}{
short=HSP,
long=high-synchrotron-peaked,
}

\DeclareAcronym{hv}{
short=HV,
long=high voltage,
}

\DeclareAcronym{iactel}{
short=IACT,
long=imaging atmospheric Cherenkov telescope
}

\DeclareAcronym{iactech}{
short=IACT,
alt=IACT technique,
long=imaging atmospheric Cherenkov technique
}

\DeclareAcronym{ibl}{
short=IBL,
long=\ac{isp} \acs*{bllac}
}

\DeclareAcronym{ic}{
short=IC,
long=inverse Compton
}

\DeclareAcronym{icm}{
short=ICM,
long=intracluster medium,
}

\DeclareAcronym{icmf}{
short=ICMF,
long=intracluster magnetic field,
}

\DeclareAcronym{igmf}{
short=IGMF,
long=intergalactic magnetic field,
}

\DeclareAcronym{integral}{
short=INTEGRAL,
long=International Gamma-Ray Astrophysics Laboratory,
cite={winkler_integral_2003},
}

\DeclareAcronym{ipac}{
short=IPAC,
long=\Acl*{ir} Processing \& Analysis Center,
}

\DeclareAcronym{ipn}{
short=IPN,
long=Interplanetary Network,
}

\DeclareAcronym{ir}{
short=IR,
long=infrared,
}

\DeclareAcronym{irf}{
short=IRF,
long=instrument response function,
}

\DeclareAcronym{irsa}{
short=IRSA,
long=\acs{nasa}/\acs{ipac} Infrared Science Archive,
}

\DeclareAcronym{ism}{
short=ISM,
long=interstellar medium,
}

\DeclareAcronym{isp}{
short=ISP,
long=intermediate-synchrotron-peaked,
}

\DeclareAcronym{itm}{
short=ITM,
long=image template method,
}

\DeclareAcronym{kait}{
short=KAIT,
long=Katzman Automatic Imaging Telescope,
cite={zheng_kait_nodate},
}

\DeclareAcronym{kaon}{
short=kaon,
long=K meson,
tag=convenience,
}

\DeclareAcronym{ksvz}{
short=KSVZ,
long = Kim-Shifman-Vainshtein-Zakharov,
}

\DeclareAcronym{lat}{
short=LAT,
long=Large Area Telescope,
cite={atwood_large_2009},
}

\DeclareAcronym{lbl}{
short=LBL,
long=\ac{lsp} \acs*{bllac}
}

\DeclareAcronym{led}{
short=LED,
long=light-emitting diode,
tag=common,
}

\DeclareAcronym{lgrb}{
short=lGRB,
long=long \acl{grb},
}

\DeclareAcronym{lhc}{
short=LHC,
long=Large Hadron Collider,
}

\DeclareAcronym{ligo}{
short=LIGO,
long=Laser Interferometer Gravitational-Wave Observatory,
cite={abbott_ligo_2009, the_ligo_scientific_collaboration_advanced_2015},
}

\DeclareAcronym{liv}{
short=LIV,
long=Lorentz invariance violation,
}

\DeclareAcronym{lp}{
short=LP,
long=log parabola,
}

\DeclareAcronym{lsp}{
short=LSP,
long=low-synchrotron-peaked,
}

\DeclareAcronym{lt}{
short=LT,
long=lookup table,
}

\DeclareAcronym{lst}{
short=LST,
long=large-sized telescope,
}

\DeclareAcronym{lsw}{
short=LSW,
long=``Light Shining through a Wall'',
}

\DeclareAcronym{mc}{
short=MC,
long=Monte Carlo,
}

\DeclareAcronym{magic}{
short=MAGIC,
long=Major Atmospheric Gamma Imaging Cherenkov Telescopes,
cite={aleksic_major_2016},
}

\DeclareAcronym{mapm}{
short=MAPM,
long=multi-anode photomultiplier,
}

\DeclareAcronym{mcmc}{
short=MCMC,
long=Markov chain \acl*{mc},
}

\DeclareAcronym{mle}{
short=MLE,
long=maximum likelihood estimation,
}

\DeclareAcronym{mou}{
short=MoU,
long=Memorandum of Understanding,
long-plural-form=Memoranda of Understanding,
}

\DeclareAcronym{mri}{
short=MRI,
long=Major Research Instrumentation,
}

\DeclareAcronym{mrk}{
short=Mrk,
long=Markarian,
}

\DeclareAcronym{msl}{
short=MSL,
long=mean-scaled length,
}

\DeclareAcronym{mst}{
short=MST,
long=medium-sized telescope,
}

\DeclareAcronym{msw}{
short=MSW,
long=mean-scaled width,
}

\DeclareAcronym{mt}{
short=MT,
long=molecular torus,
}

\DeclareAcronym{mw}{
short=MW,
long=Milky Way,
}

\DeclareAcronym{mwl}{
short=MWL,
long=multi-wavelength
}

\DeclareAcronym{nasa}{
short=NASA,
long=National Aeronautics and Space Administration,
}

\DeclareAcronym{ned}{
short=NED,
long=\acs*{nasa}/\acs*{ipac} Extragalactic Database,
}

\DeclareAcronym{nd}{
short=ND,
long=neutral density,
}

\DeclareAcronym{ndaf}{
short=NDAF,
long=neutrino-dominated accretion ﬂow,
}

\DeclareAcronym{nlr}{
short=NLR,
long=narrow-line region,
}

\DeclareAcronym{ns}{
short=NS,
long=neutron star,
}

\DeclareAcronym{nsbh}{
short=NSBH,
long=\acl{ns} \acl{bh},
}

\DeclareAcronym{nsb}{
short=NSB,
long=night sky background,
}

\DeclareAcronym{nsf}{
short=NSF,
long=National Science Foundation,
}

\DeclareAcronym{nu-antinu}{
short=\ensuremath{\nu \overline{\nu}},
long=neutrino/anti-neutrino,
tag=convenience,
}

\DeclareAcronym{od}{
short=OD,
long=optical density,
}

\DeclareAcronym{oss}{
short=OSS,
long=optical support structure,
}

\DeclareAcronym{ovro}{
short=OVRO,
long=Owens Valley Radio Observatory,
cite={richards_blazars_2011},
}

\DeclareAcronym{pc}{
short=pc,
short-plural=,
long=parsec,
tag=unit,
}

\DeclareAcronym{pcb}{
short=PCB,
long=printed circuit board,
}

\DeclareAcronym{pe}{
short=p.e.,
long=photoelectron,
}

\DeclareAcronym{pion}{
short=pion,
long=pi meson,
tag=convenience,
}

\DeclareAcronym{pmt}{
short=PMT,
long=photomultiplier tube,
}

\DeclareAcronym{png}{
short=pNG,
long=pseudo Nambu-Goldstone,
}

\DeclareAcronym{pq}{
short=PQ,
long=Peccei-Quinn,
}

\DeclareAcronym{pqww}{
short=PQWW,
long=Peccei-Quinn-Weinberg-Wilczek,
}

\DeclareAcronym{psct}{
short=pSCT,
sort=pSCT,
long=prototype \acl{sct},
}

\DeclareAcronym{psf}{
short=PSF,
long=point spread function,
}

\DeclareAcronym{psoc}{
short=PSoC,
long=programmable system on a chip,
}

\DeclareAcronym{psu}{
short=PSU,
long=power supply unit,
}

\DeclareAcronym{pwn}{
short=PWN,
short-plural=e,
long=pulsar wind nebula,
long-plural-form=pulsar wind nebulae,
}

\DeclareAcronym{qcd}{
short=QCD,
long=quantum chromodynamics,
}

\DeclareAcronym{qe}{
short=QE,
long=quantum efficiency,
}

\DeclareAcronym{qed}{
short=QED,
long=quantum electrodynamics,
}

\DeclareAcronym{ra}{
short=RA,
long=right ascension,
}

\DeclareAcronym{ras}{
short=RAS syndrome,
long=recurrent acronym syndrome,
}

\DeclareAcronym{rbm}{
short=RBM,
long=ring background method,
}

\DeclareAcronym{rg}{
short=RG,
long=radio galaxy,
long-plural-form=radio galaxies,
}

\DeclareAcronym{rm}{
short=RM,
long=rotation measure,
}

\DeclareAcronym{rms}{
short=RMS,
long=root mean square,
}

\DeclareAcronym{roi}{
short=RoI,
long=region of interest,
}

\DeclareAcronym{rr}{
short=RR,
long=reflected region,
}

\DeclareAcronym{sc}{
short=SC,
long=Schwarzschild-Couder,
}

\DeclareAcronym{sct}{
short=SCT,
long=\acl{sc} telescope,
}

\DeclareAcronym{sdss}{
short=SDSS,
long=Sloan Digital Sky Survey,
}

\DeclareAcronym{sed}{
short=SED,
long=spectral energy distribution,
}

\DeclareAcronym{sgrb}{
short=sGRB,
long=short \acl{grb},
}

\DeclareAcronym{sipm}{
short=SiPM,
long=silicon photomultiplier,
}

\DeclareAcronym{msun}{
short=\ensuremath{{\rm M}_{\odot}},
sort=SM,
short-plural=,
long=solar mass,
long-plural-form=solar masses,
tag=unit,
}

\DeclareAcronym{sm}{
short=SM,
long=Standard Model,
}

\DeclareAcronym{smbh}{
short=SMBH,
long=supermassive black hole,
}

\DeclareAcronym{smns}{
short=SMNS,
long=supra-massive \acs*{ns},
}

\DeclareAcronym{spe}{
short=SPE,
long=single \acl*{pe},
}

\DeclareAcronym{ssc}{
short=SSC,
long=synchrotron self-Compton
}

\DeclareAcronym{sst}{
short=SST,
long=small-sized telescope,
}

\DeclareAcronym{svd}{
short=SVD,
long=singular-value decomposition,
}

\DeclareAcronym{svo}{
short=SVO,
long=Spanish Virtual Observatory,
}

\DeclareAcronym{s/n}{
short=S/N,
long=signal-to-noise ratio,
first-style = long,
subsequent-style = long,
}

\DeclareAcronym{ttl}{
short=TTL,
long=transistor-transistor logic,
}

\DeclareAcronym{ul}{
short=UL,
long=upper limit,
}

\DeclareAcronym{usb}{
short=USB,
long=universal serial bus,
tag=common,
}

\DeclareAcronym{uv}{
short=UV,
long=ultraviolet,
}

\DeclareAcronym{uvot}{
short=UVOT,
long=\Acs*{uv}/Optical Telescope,
cite={roming_first_2009},
}

\DeclareAcronym{vbf}{
short=VBF,
long=\acs*{vts} Bank Format,
}

\DeclareAcronym{vev}{
short=VEV,
long=vacuum expectation value,
}

\DeclareAcronym{vhe}{
short=VHE,
long=very-high energy,
}

\DeclareAcronym{vegas}{
short=\pkg{VEGAS},
sort=VEGAS,
long=\acs*{vts} Gamma-ray Analysis Suite,
}

\DeclareAcronym{vlba}{
short=VLBA,
long=Very Long Baseline Array,
}

\DeclareAcronym{vts}{
short=VERITAS,
long=Very Energetic Radiation Imaging Telescope Array System
}

\DeclareAcronym{wdr}{
short=WDR,
long=whole dish reflectivity,
}

\DeclareAcronym{wimp}{
short=WIMP,
long=weakly interacting massive particle,
}

\DeclareAcronym{wisp}{
short=WISP,
long=weakly interacting sub-eV particle,
}

\DeclareAcronym{xrt}{
short=XRT,
long=X-ray Telescope,
cite={burrows_swift_2005}
}